\documentclass[%
reprint,
amsmath,amssymb,
aps,
prl,
floatfix,
preprintnumbers
]{revtex4-2}

\usepackage{graphicx}
\usepackage{dcolumn}
\usepackage{bm}
\usepackage{natbib} 
\usepackage{booktabs} 
\usepackage{placeins}

\newcommand{\thetabf}{\bm{\theta}}
\newcommand{\hbf}{\bm{h}}
\newcommand{\sbf}{\bm{s}}
\newcommand{\nbf}{\bm{n}}
\newcommand{\dbf}{\bm{d}}
\newcommand{\Cbf}{\bm{C}}
\usepackage{lineno}
\begin{document}
\preprint{LIGO-P2600015}

\title{Multi-Segment Consistency Tests of General Relativity}

\author{Vaishak Prasad}
\email[]{vaishakprasad@psu.edu}
\affiliation{Institute for Gravitation and the Cosmos, Department of Physics, Penn State University, University Park, Pennsylvania, 16801, USA}
\affiliation{Department of Astronomy and Astrophysics, Penn State University, University Park, Pennsylvania, 16801, USA}

\date{\today}

\begin{abstract}
As the LIGO-VIRGO-KAGRA Network of gravitational-wave detectors improves in sensitivity, accumulating hundreds of gravitational-wave detections per year, it becomes imperative to improve tests of general relativity in concert. The test of Hawking's law of area increase has gained prominence since GW250114, where black holes in General Relativity were tested with unprecedented precision, using the linear ringdown and pre-merger portions of the signal. A closely related test is the Inspiral-Merger-Ringdown Consistency Test, which assesses the consistency of the high- and low-frequency parts of the signals. In this letter, I present a multi-parameter Multi-Segment Consistency Test (MSCT) that generalizes and improves upon existing tests by ensuring that the extrinsic properties of the signal are consistent across its independent segments and by adopting an accelerated time-domain approach. The improved area law test is then presented as a projection of this MSCT test. These crucial improvements, which bring physical consistency to the area law test, lead to more stringent constraints on the increase in estimated area from observed binary black hole mergers, while also capturing covariances among the parameters. 

Applying the two-segment version of this test to the inspiral and ringdown parts of GW250114, and keeping some of the extrinsic parameters common between the segments, I test the signal to unprecedented accuracy, obtaining $4.61 ^{+0.24} _{-0.11}\sigma$ significant result for the area increase, even as more than 4 pre-merger cycles of the signal are excluded from the analysis. Also, I infer that the final state lies within the 15\% highest posterior density confidence interval.

\end{abstract}

\maketitle


\paragraph{Introduction}

Ever since the first detection of gravitational waves with GW150914 \cite{GW150914}, improvements in sensitivity (see e.g. \cite{Abbott2016_AdvancedLIGO_FirstDiscoveries,Martynov2016_SensitivityBeginning,Tse2019_QuantumEnhancedALIGO,Acernese2019_VirgoSqueezing,Buikema2020_ALIGO_O3Performance,Capote2025_ALIGO_O4Performance,VirgoCollaboration2025_AdV_O4OpticalCharacterization,KAGRACollaboration2019_NatAstron,Akutsu2021_KAGRA_DesignHistory,Cooper2023_Aplus_SensorsActuators, LIGOScientific:2025hdt}), detection methodologies \cite{Messick2017_GstLAL_PromptDiscovery,Nitz2018_PyCBCLive,Davies2020_ExtendingPyCBC,Davis2022_PyCBC_DataQualityStreams,Tsukada2023_GstLAL_RankingStats,Cannon2021_GstLAL_SoftwareX,SingerPrice2016_BAYESTAR,Chaudhary2024_LowLatencyAlertProducts_O4,Drago2021_cWB_SoftwareX,Mishra2022_cWB_ML_BBH_O3,Szczepanczyk2023_Burst_O3_cWB_ML,Cornish2021_BayesWave_EraOfObservations,Zevin2017_GravitySpy,Coughlin2019_NovelGlitches_DeepTransferLearning}, and waveform modeling \cite{Husa2016_IMRPhenomD_Base,Khan2016_IMRPhenomD_Model,Hannam2014_PhenomP,Schmidt2015_ChiP_SingleSpinPrecession,London2018_PhenomHM,Garcia-Quiros2020_IMRPhenomXHM,Khan2020_PhenomPv3HM,Pratten2021_IMRPhenomXPHM,Yu2023_IMRPhenomXODE,BuonannoDamour1999_EOB,BuonannoDamour2000_Plunge,Taracchini2014_SEOBNRv2,Purrer2016_SEOBNRv2_ROM,Bohe2017_SEOBNRv4,Cotesta2018_SEOBNRv4HM,Ossokine2020_SEOBNRv4PHM,Ramos-Buades2023_SEOBNRv5PHM,Babak2017_PrecessingEOB_Validation,Varma2019_NRHybSur3dq8,Varma2019_NRSur7dq4,FlanaganHinderer2008_TidalLove,Vines2011_Post1PN_Tides,Dietrich:2017aum,Dietrich2019_MatterImprints,Dietrich2019_ImprovedNRTidal,Nagar2018_TEOBResumS,Akcay2021_TEOBResumSP,Cao2017_SEOBNRE,Chiaramello2020_EOB_Eccentric,Paul2025_ESIGMAHM} and systematics \cite{Babak2017_PrecessingEOB_Validation,Cotesta2018_SEOBNRv4HM,Purrer2016_SEOBNRv2_ROM,Varma2019_NRHybSur3dq8,Varma2019_NRSur7dq4,Ossokine2020_SEOBNRv4PHM,Ramos-Buades2023_SEOBNRv5PHM,Garcia-Quiros2020_IMRPhenomXHM,Khan2020_PhenomPv3HM,Pratten2021_IMRPhenomXPHM, Colleoni:2024knd,Dietrich:2017aum,Dietrich2019_MatterImprints,Dietrich2019_ImprovedNRTidal,Nagar2018_TEOBResumS,Akcay2021_TEOBResumSP,Chiaramello2020_EOB_Eccentric} have led to more than 400 potential events, unto the fourth observing run alone. The early detections of events, which were detected at SNRs of around 10 - 40, helped validate the detection pipelines and the early efforts that modeled the dominant non-spinning quasi-circular effects in waveforms using phenomenological, effective one-body and surrogate methods, enabling the construction of large banks of templates in a limited time, and the adaptation of efficient matched filtering techniques to detect gravitational waves.  This helped establish a solid foundation that has led to continuous incremental improvements over the past decade, with waveform models in the current generation being able to model complex signal morphology, including some post-Newtonian effects up upto 5.5PN order, spin-precession effects, and asymmetric masses. Owing largely to these improvements, the detection of an increasing number of high-signal-to-noise signals~\cite{LIGOScientific:2025cmm, LIGOScientific:2025brd, LIGOScientific:2025rid, LIGOScientific:2025obp} and improvements in data analysis strategies, testing general relativity has become a central focus. 

Currently, there is a plethora of tests aimed at evaluating various features of signals. This includes the chi-squared test aimed at testing the power distribution of the signal across frequency bins \cite{Allen2005_Chi2TF}, dispersion relation \cite{Mirshekari2012_ModDispersion, baka2025testinggeneralrelativitygravitational} and polarization tests \cite{Nishizawa2009_Polarizations, wong2021nullstreambasedbayesianunmodeledframework} that test the propagation/polarization properties of the signal, the Inspiral Merger Ringdown consistency test (IMRCT) \cite{Ghosh2016_IMRConsistency, Ghosh:2017gfp}  that aim to test the consistency between low and high and low frequency parts of the signal, and the ringdown/ black hole spectroscopy tests \cite{Gossan2012_NoHair_Bayes, Carullo:2019flw, Forteza:2023APConsistency} that explicitly tests the linear quasi-normal mode structure from the post-merger part of the signal. Other tests, such as the pSEOB \cite{Maggio:2023pPMRD, Pompili:2025cdc}, modify the merger-ringdown portion by varying the quasi-normal mode structure to detect deviations in the ringdown phase. There are further Post-Einsteinian and Post-Newtonian tests that vary the amplitudes and phases of the waveforms, or the post-Newtonian coefficients in the inspiral phase \cite{YunesPretorius2009_ppE, Cornish2011_ppE_Bayes, DelPozzo2011_TIGER, Agathos2014_TIGERpipeline, Arun2006_PNcoeffTest, tiger_23, fti_26}, to constrain deviations from GR, to name a few. See \cite{Abbott2016_TGR_GW150914, Abbott2019_GWTC1_GRTests, Abbott2021_TGR_GWTC2, Abbott2025_TGR_GWTC3, LIGOScientific:2017vwq} for a non-exhaustive overview. 

Of particular interest to this work is the IMRCT, 
which splits the signal into a low-frequency inspiral part and a high-frequency post-inspiral band, and analyzes them independently, assuming GR in the frequency domain. The posterior samples for the final remnant black hole's mass and spin are obtained from each and compared. The IMRCT is a powerful test that has been used profusely to test GR across events from multiple observing runs. The IMRCT and other tests of GR have also been used to study the biases in null tests of GR induced by certain physical effects such as precession~\cite{Hamilton:2023znn}, eccentricity~\cite{Bhat:2022amc, Narayan:2023vhm, Shaikh:2024wyn}, and lensing~\cite{Narayan:2024rat}.


Another test of interest here is the test of Hawking's area theorem \cite{Hawking1971_AreaTheorem}, which states that, assuming general relativity and the null energy condition, the area of black hole horizons can only increase in physical processes. In the context of binary black hole mergers, this means that the late-time area of the remnant black hole should be larger than the sum of the initial areas of the progenitors. 

The gravitational signal from a binary black hole merger consists roughly of three phases: the inspiral phase, where the progenitor black holes are inspiralling at a distance, with their orbits slowly shrinking due to the loss of energy and angular momentum to gravitational radiation, (ii) the plunge-merger phase, when no more stable circular orbits exist, leading to their pluge towars each other. This leads to the formation of a common dynamical horizon that envelopes the two individual horizons, and the amplitude of the gravitational radiation is maximum around this time \cite{Cadez1974_CommonAH, CookAbrahams1992_HorizonStructure, Anninos1995_HeadOnCommonAH, Gupta:2018znn, Prasad:2023bwa}. (iii) the post-merger phase where the formed common dynamical horizon settles down to a stationary Kerr horizon by absorbing gravitational radiation, with the spacetime emitting gravitational radiation in the process.

The gravitational radiation from the early-time inspiralling phase of a BBH evolution can be efficiently described by post-Newtonian modelling \cite{Living:Blanchet}, whereas the late-time post-merger gravitational waveform in General Relativity is described by a superposition of quasi-normal modes and later-time power-law tails \cite{Vishveshwara:1970zz, Price:1971fb, Teukolsky:1973ap, Teukolsky:1972my, Berti:2009kk}. However, to accurately describe the most dynamical non-linear plunge-merger phase of BBH mergers, one needs an accurate waveform from numerical relativity simulations \cite{Pretorius:2005gq, Campanelli:2005dd, Scheel:2025jct, Mroue:2013, Boyle:2019SXS, Jani:2016GT,Healy:2017RIT, Healy:2019RIT2, Healy:2020RIT3, Ferguson:2025MAYA2, Dietrich:2017aum, Gonzalez:2023CoRe2}. It is this phase of a BH merger that leads to the emission of a significant fraction of gravitational radiation and contributes to the largest increase in area \cite{prasad_area}. Testing this phase of BBH mergers amounts to testing the most nonlinear and dynamical features of General Relativity. Hence, a scheme that quantitatively tests the area increase in the binary merger scenario is crucial to testing infrastructure.

The analysis of GW signals and the parameter estimation of GW sources are currently carried out using a Bayesian framework \cite{Finn:1992wt}. The posterior distribution of the model parameters describing the signal is given by:
\begin{equation}
    p(\thetabf | \dbf) = \dfrac{\mathcal{L}(\dbf | \thetabf) \pi(\thetabf)}{z} \label{eqn:bayes}
\end{equation}
Where $\mathcal{L}(\dbf | \thetabf)$ is the likelihood function for observing the data under the model parameters, $\pi(\thetabf)$ is the prior information on the model parameters, and $z=\int \mathcal{L}(\dbf | \thetabf) \pi(\thetabf)d\thetabf$ . The analysis and the Likelihood function are constructed in the frequency domain \cite{Finn:1992xs, Cutler:1994ys}, assuming a wide-sense stationary Gaussian random process for the noise model. If the domain of analysis is periodic, then the noise covariance matrix is circulant and diagonal in the Fourier basis. All the properties of the zero-mean noise random process are then completely described by the single-sided Power spectral density (PSD) $S_n(f)$, and an inner product using the matched filter \cite{Turin1960_MatchedFilters, Helstrom1968_StatisticalTheory, WainsteinZubakov1970_Extraction, Sathyaprakash:1991mt, Dhurandhar:1992mw} can be defined:
\begin{equation}
    \langle \tilde{\hbf}, \tilde{\sbf} \rangle_{FD} = 4 Re \left( \int_0 ^\infty{\dfrac{\tilde{h}(f) \tilde{s}^*(f) df}{S_n(f)}}\right) \label{mf}
\end{equation}
Given the single sided data $\tilde{d(f)}$, a signal $\tilde{s}(f)$, a template $\tilde{h}(f)$ that aims to model $\tilde{s}(f)$, and noise $\tilde{n}(f)$ that adds linearly to the signal given by $\tilde{\dbf} = \tilde{\sbf} + \tilde{\nbf}$, one can define the Gaussian Likelihood Model:
\begin{equation}
    \log \mathcal{L}_{FD}(\dbf | \thetabf) = -\dfrac{1}{2}\langle (\tilde{\dbf}-\tilde{\hbf}), (\tilde{\dbf}-\tilde{\hbf})\rangle_{FD} \label{fdl}
\end{equation}

Testing Hawking's area theorem amounts to verifying that the total area indeed increases in a binary merger scenario.  The increase in the area of the horizons in the due process of a BBH merger was tested for the first time using an injected GW150914 signal \cite{Cabero:2017avf}. This involved analyzing the pre-merger and post-merger portions of the signal separately in the frequency domain and time domain, respectively. The pre-merger portion of the signal is appropriately tapered before analyzing it in the frequency domain. The post-merger signal is analyzed on a grid of start times, and the Jensen–Shannon divergence is used to identify a preferred start time at which inference stabilizes. The initial areas are determined from the inspiral analysis, and the final remnant areas are determined from the dominant QNMs analysis of the post-merger signal. 

In \cite{Isi2021AreaLaw}, which applied it to real data, a time-domain analysis strategy is adapted to both parts of the signal. The entire Inspiral-Merger-Ringdown (IMR) signal is split at the peak of the signal, and the pre-peak and post-peak parts of the signal are analyzed separately in the time domain. The posterior samples from the pre-peak analysis are used to compute the initial areas, while the QNM analysis of the post-peak signal is used to determine the remnant areas. By taking a product of the posterior samples from the inspiral and ringdown analyses, given their assumed independence, one can arrive at the posterior distribution for the increase in area:
\begin{equation}
    p(\Delta A | \dbf) = P(A_f - A_i | (A_i, A_f) \in \{p(A_i)|\dbf\} \times \{p(A_f | \dbf)\} ) \label{eqn:post_ainc}
\end{equation}

In a recent work, the GW250114 signal from a BBH merger, recorded with an SNR of 76 in the LIGO network of detectors, was used to test Hawking's area increase law \cite{LIGOScientific:2025rid, LIGOScientific:2025obp, Miller:tdinf:2025}. The spirit of the method remained the same, but the inspiral-phase analysis was terminated at $-40M$ before the peak, thereby excluding the two loudest signal cycles near the merger. 

One important problem that persists in existing consistency tests and area-law analyses is that, although the pre- and post-merger parts of the signals originate from the same source, they are treated as entirely independent in the analysis. This is because each sample from the inspiral (ringdown) analysis is a point in the 15-dimensional prior space, and completely determines the entire signal, including the extrinsic parameters of the source, i.e., the luminosity distance $d_L$, inclination angle $\iota$ (or equivalently the angle between the total angular momentum of the source and the line of sight $\thetabf_{JN}$), time of arrival $t_{geo}$, the right ascension $\alpha$, and declination $\delta$. Even if the post-inspiral signal does not follow the predictions of general relativity, the intrinsic parameters of the signal would change (but would be mapped to general relativity in the test). However, some of the extrinsic parameters arguably would not, as the different parts of the signal come from the same source. By taking a product of posterior samples from independent analyses of the different parts of the signal, one ignores this crucial single-source constraint, including signals described by inconsistent extrinsic parameters across the two segments. One side effect of this would be that the posteriors for the area increase would be broader than expected.

Previously, \cite{Cabero:2017avf} suggested an ad hoc solution to this problem. They suggest using multiple MCMC walkers, each with fixed extrinsic parameters drawn from the 50\% confidence interval of the posterior distribution derived from the analysis of the full IMR signal. By using the same set of sampled extrinsic-parameter-fixed walkers across the inspiral and ringdown analyses, they propose to address this issue.

However, a truly consistent and robust area law analysis should also sample over extrinsic parameters, allowing correlations between extrinsic and intrinsic parameters. The extrinsic parameters that determine the location of the source and its spatial orientation are purely geometric and independent of the underlying physical theory. Therefore, when testing for consistency between two portions of the signal, one must acknowledge this. The first step in this direction was taken by \cite{CorreiaCapano:2024SkyMarginalization, Correia2}, where the sky location $\alpha, \delta$ and the peak time $t_c$ are sampled over but kept common between the pre-merger and post-merger parts of the signal. Their analysis was implemented in the frequency domain using gating and inpainting techniques~\cite {Wang:2023DataGapsTianQin,Biwer:2018osg,CorreiaCapano:2024SkyMarginalization}. A similar method was applied to analyze the area increase in GW230814 more recently in \cite{Tang:2025GW230814AreaLaw}.

The gating and inpainting method, which provides a way to down-weight unused parts of a signal, enables one to analyze sufficiently well-localized time segments of the signal while enjoying the efficiency of a frequency domain analysis. Note that if not for the gating and inpainting technique, before the accelerations introduced in~\cite{prasad_tda}, it was impossible to analyze limited time segments efficiently.
But, the technique can introduce instabilities due to edge effects \cite{Harris1978WindowsDFT} if continuity is not enforced across the gated boundaries.

Since Fourier-domain likelihoods assume approximate circulancy of the noise covariance matrix, gating is typically accompanied by tapering to suppress spectral leakage and mitigate Gibbs-type oscillations \cite{Dahlhaus1988SmallSample, Guillaumin2022DebiasedWhittle, Wang2023DataGapsTianQin, Pilz2012TaperingWindowed}. The accuracy of the analysis, therefore, depends on both the gating configuration and the tapering window. Incomplete removal of the inspiral portion or leakage of the taper into the ringdown segment can bias the inferred post-merger signal. Given the additional ambiguity in defining the ringdown start time, the resulting inference is intrinsically sensitive to these choices.


This work aims to fully resolve these issues and improve upon previous efforts by casting the analysis entirely in the time domain, thereby providing an independent analysis pipeline. In this letter, I propose a robust approach to test GR by mapping the signal's parameters in different segments to GR, also ensuring complete consistency of the analysis across the multiple independent analysis segments, and efficiency using an accelerated time-domain analysis, enabling clear temporal demarcations. I apply this to the celebrated GW250114 event, demonstrating its efficacy and strengths in testing the consistency of different parts of the signals, thereby enabling stronger tests of the highly nonlinear dynamical phases of general relativity. Adopting the suggestion to quantify the area increase in \cite{prasad_area}, I show how one can robustly test not only for area increase but also quantify deviations from GR during the process. The end result is a cross segment consistency test of GR that can be applied to any number of time limited portions of the signals, allowing one to consistently infer the parameters of the signal in each individual segment, while at the same time adhering to the constraint that they are coming from the same source, without the need for the periodic domain assumption, or tapering.

\paragraph{Methodology}

To test the area increase during the due process of the merger of a BBH system, I use the time-domain representation of the signal as implemented in \emph{tdanalysis}~\cite{prasad_tda}, similar to \cite{Isi2021AreaLaw, Miller:tdinf:2025}. This ensures a clean separation of the inspiral and ringdown phases of the signals, since mere filtering in the frequency domain does not guarantee time localization.  In contrast to the existing pipelines, \emph{tdanalysis} is highly optimized, enabling time-domain analysis and stochastic sampling of high-dimensional posteriors, without the need to fix certain parameters, in a feasible time. 

The inspiral signal would be sharply cutoff at a fixed time $t_{I}$ before the peak, and the ringdown signal would sharply begin at a fixed time $t_R$. No windows or tapering would be necessary either at the end of the inspiral or the beginning of the ringdown. A crucial difference between existing approaches to analyzing the post-merger signal is that, instead of spectroscopic analysis of the ringdown phase, I directly use waveform models to model the post-merger waveform.


The analysis then consists of these two semi-independent portions, decided in the signal by sharp cutoffs in the time domain.  

I also adopt a Bayesian approach to obtain the posterior distribution of the model parameters jointly describing these two parts of the signal. This is achieved by defining a likelihood function that effectively encodes the assumption of wide-sense stationary Gaussian noise in the time domain, without assuming periodicity in the domain. The inner product in the time domain is then written as:
\begin{equation}
   \langle \hbf, \sbf \rangle = \hbf^T \Cbf^{-1} \sbf \label{eqn:ip_td}
\end{equation}
With $\Cbf$ being a symmetric Toeplitz matrix describing the second moment of the noise.  The specific details of the implementation of the accelerated time-domain code \emph{tdanalysis} are presented in ~\cite{prasad_tda}. The reader may also refer to \cite{Kalman1960KalmanFilter,Durbin1960FittingTimeSeries, BoxJenkins1970TimeSeries,isi2021analyzingblackholeringdowns} for general details.

The waveforms are described by a preferred waveform model, in this letter chosen to be NRSur7dq4~\cite{Varma2019_NRSur7dq4}. 

The inspiral part of the signal is described using 8 intrinsic parameters (the component masses $M_{1,2}$ and their spin vectors $\vec{S}_{1,2}$) and 7 extrinsic parameters (the luminosity distance $d_L$, the right ascension $\alpha$, declination $\delta$, the reference phase $\phi_{ref}$, polarization angle $\psi$, source inclination $\iota$ or $\theta_{JN}$ and the GPS time of the signal peak $t_{gps}$). This leads to a likelihood function for the inspiral portion:
\begin{equation}
    \log \mathcal{L}_{I} (\dbf | \thetabf_{insp}) = -\dfrac{1}{2}\langle d(t) - h(t| \thetabf_{insp}),   d(t) - h(t | \thetabf_{insp})\rangle \label{eqn:insp_lik}
\end{equation}
Where $\thetabf_i$ are the 15-dimensional inspiral parameters of the system.

The ringdown portion of the signal is also described by the corresponding post-merger portion of a BBH waveform from the same waveform model, which is chosen here to be NRSur7dq4, and therefore requires 15 parameters to specify. One major advantage is that it includes all angular modes and overtones used to model the waveform. A second advantage is that the start of the template waveform is automatically determined by the inference procedure, thereby eliminating ambiguities, as in spectroscopic analysis. I refer the reader to ~\cite{prasad_tda} for a discussion on the differences of this implementation with the existing works. 


The framework used here is also an improvement over the existing frameworks in several aspects. First, as is the case in ~\cite{Miller:tdinf:2025}, it is a fully time domain pipeline with a Gaussian Likelihood, thus not needing the Whittle approximation. Second, owing to the major acceleration efforts reported in ~\cite{prasad_tda}, all parameters are sampled over, including the sky locations and the peak time. These, along with sometimes the polarization angle, have generally been fixed for speed in past works~\cite{CalderonBustillo:2020rmh,Qiu:2023, Gennari:2023gmx,Chandra:2025, Miller:2023ncs, Miller_2025}. 

Third, once the analysis segments are defined by their GPS times, the peak time of the signal determines the portion of the template waveform that lies in the analysis window, including for the ringdown portion, thus dynamically determining it in the stochastic sampling process. This naturally allows relevant portions of the waveforms to be used to model the corresponding part of the signals. In the post-merger segment, this enables the use of approximants that model the non-linear portions of the waveform by calibration to NR, and include higher-order modes.
In the case of NRSur7dq4, these include most of the angular modes and overtones resolved by current numerical relativity simulations in the SXS catalog \cite{Scheel:2025jct}.

Post-merger portions of waveforms have been used in the past to model the post-peak part of the signal, however with certain parameters fixed~\cite{CalderonBustillo:2020rmh,Gennari:2023gmx, Chandra:2025}. In this analysis, no parameters are fixed. The peak-time, which is shared between the inspiral and post-merger analysis segments, is also a sampled parameter.  This removes the linear ringdown start time ambiguities, also enabling one to effectively marginalize over the start time of the waveform. Thus, one is not expected the analysis and tests of GR, such as the area law analysis, to be impacted by linear-ringdown modeling systematics as demonstrated in the case of GW250114~\cite{prasad_area}.

Without the acceleration techniques of ~\cite{prasad_tda}, it would be impossible to carry out simultaneous high-dimensional inference across the segments analyzed here. 

 The likelihood model for the ringdown segment can be written as:
\begin{equation}
    \log \mathcal{L}_{R} (\dbf | \thetabf_{rd}) = -\dfrac{1}{2}\langle d(t) - h(t| \thetabf_{rd}),   d(t) - h(t | \thetabf_{rd})\rangle \label{eqn:rd_lik}
\end{equation}
However, in order to impose consistency between the extrinsic parameters of the different parts of the signal, we do not consider the parameter sets $\thetabf_{insp}$ and $\thetabf_{rd}$ to be fully exclusive. I separate out certain extrinsic parameters from each set and make them common between the two disjoint segments in the time domain, leading to a likelihood model of the form:
\begin{equation}
     \log \mathcal{L}(\dbf | \thetabf) = \log \mathcal{L}_{I} (\dbf | \thetabf_I, \thetabf_c ) +   \log \mathcal{L}_{R} (\dbf | \thetabf_R, \thetabf_c) \label{eqn:lik_td}
\end{equation}
Where $\thetabf_{insp} \equiv \thetabf_I \cup \thetabf_c $ and  $\thetabf_{rd} \equiv \thetabf_R \cup \thetabf_c $, with $\thetabf_I$ and $\thetabf_R$ describing the exclusive inspiral and ringdown parameters of the system, and $\thetabf_c$ being the common set of parameters across the segments. In this example, I keep the parameters $d_L, \alpha, \delta, t_{c}$, and $\thetabf_{JN}$ fixed, although our framework is flexible. In ~\ref{sec:ss}, I provide our rationale for this default choice of commonly sampled parameters.

Thus, each proposal in the prior space corresponds to a 25-dimensional parameter vector, with 10 exclusive parameters each describing the inspiral and post-merger signal, and 5 parameters that are also sampled over, but common across the models. The waveform is jointly described by generating it with the $\thetabf_{insp}$ and $\thetabf_{rd}$ parameters and truncating it at predetermined times.

The bilby parameter estimation infrastructure with the dynesty sampling backend is used to integrate the evidence and obtain posterior samples via nested sampling. Waveforms are generated in the time domain by a direct call to lalsimulation.

\paragraph{The Area test}
The resultant posterior samples span a 25-dimensional parameter space, and contain the set of parameters \{$M_{1,2}, \vec{S}_{1,2}, \psi, \phi_{ref}$\} for each considered waveform segment, and the common set of extrinsic parameters \{$d_L, \alpha, \delta, t_{gps}$\} for each sample. 

Given the set of posterior samples, one can directly test the consistency of the $n$ (2) time segments by forming the set of posterior samples obtained by subtracting the respective exclusive parameters:
\begin{equation}
 p(\Delta \thetabf | \dbf) = p(\thetabf_{I} - \thetabf_R | \dbf) \label{eqn: post_diff}
\end{equation}
This is possible because each step the sampler takes in the 25-dimensional prior space determines both the inspiral and ringdown segments, thereby analyzing them simultaneously. The semi-independent decomposition of the likelihood in \eqref{eqn:lik_td} ensures that the analysis is consistent with a single source. There is flexibility in the formulation to choose which of the parameters should be common, but for the purposes of demonstration, I stick to the 25-dimensional decomposition described above. 

There is no constraint on how long or where the two time segments cen lie in the analysis. But an interesting scenario is to delete the part of the signal corresponding to the most dynamical and nonlinear phase of BBH mergers, i.e., the late inspiral-merger phase. If the most dynamical nonlinear regime of the signal, which is excluded from the analysis, is described by GR, the parameters estimated from the post-merger and early inspiral segments of the signal must be consistent with each other, which is a full consistency test in 10 dimensions (or chosen number of exclusive parameters). 

While a full measure of agreement between the two signal segments can be obtained from the multi-dimensional posteriors above, one can also project the multi-dimensional posterior distribution onto a lower (e.g., one-dimensional) summary estimator of interest. Consider an estimator $\hat{X}_{seg} = \hat{X}(\thetabf_{seg}, \thetabf_c)$ where $\thetabf_{seg}$ denotes the exclusive parameters of the segment over which the estimator $\hat{X}$ is being computed. Then, the difference estimator $\Delta \hat{X} = \hat{X}_{R} - \hat{X}_I$ is a useful descriptor for measuring the inconsistency of the property described by $\hat{X}$ across the segments:
\begin{equation}
    E[\Delta f(\hat{X})] = \sum_j w_j ( f_{R}(\hat{X}(\thetabf_{rd, j}) - f_{I}(\hat{X}(\thetabf_{insp, j})) \label{eqn: exp_stat}
\end{equation}
Where $w_j$ are the normalized importance weights. 

If the quantity $\hat{X}$ is chosen to be the total area of horizons of black holes at the asymptotic ends of each segment, then the projection becomes identical to the area law test, with an additional improvement that the signals in the two segments are consistent with a single source. If the set of common parameters is null i.e. $\{\theta_{insp} \cap \theta_{rd}\} = \emptyset $, then the posterior samples from the two segments are independent and would be similar to the test carried out in \cite{LIGOScientific:2025rid}, with the ringdown analysis being waveform model based instead of spectroscopy based.

\paragraph{Results}
\begin{figure}
    \centering
    \includegraphics[width=\linewidth]{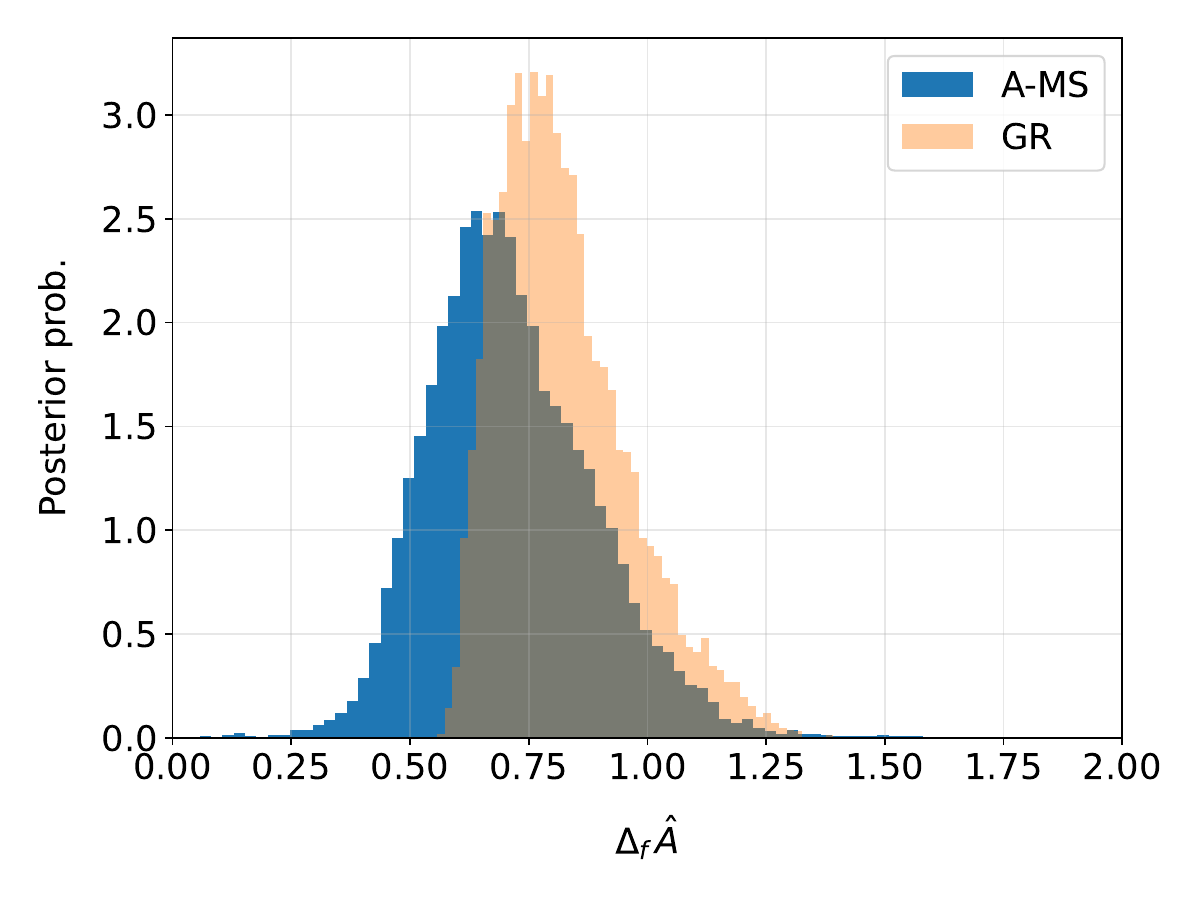}
    \caption{The fractional area increase from the MSCT, carried out on GW250114. The blue histogram shows the area increase from the multi-segment method, while that in orange shows the expected GR area increase from the posterior samples obtained from the full IMR PE.}
    \label{fig:ainc}
\end{figure}
The analysis was run using an inspiral segment end time of $-200M$ before the peak, as determined by the maximum-likelihood value for the peak in the full-time-domain IMR analysis. The ringdown segment was chosen to start at $10M$ after the peak and end at $+200M$. The matched filter network SNR of the pre-merger signal was $\approx 29.5$, while that of the ringdown was $27.4$. The combined SNR of the two segments across the detectors was around 40. The segment configuration was chosen to approximately have the same SNR in the pre- and post-merger parts of the signal. 

The resultant posterior samples contain the set of parameters  \{$M_{1,2}, \vec{S}_{1,2}, \psi, \phi_{ref}$\} for each considered waveform segment, and the common set of extrinsic parameters \{$d_L, \alpha, \delta, t_{gps}$\} for each sample. I compute the initial areas using the masses and spins from the inspiral set, and the final areas using the fitting formulae/predictions for the remnant's final masses and spins, given the post-merger waveform parameters using the EOB implementation. This can also be done using the NR ringdown surrogate. The difference between the areas computed using these parameters and those obtained from the Kerr formula yields the posterior distribution for the area increase (or, alternatively, the fractional area increase), shown in Fig.~\ref{fig:ainc}. 
 
To test for an increase in area, I consider the null hypothesis $H_0$: The final area is larger than the sum of the initial areas, and compute the probability that $H_0$ is true. This is done by using the nested (importance samples) and computing $\hat{p}_{H_1} = p(\thetabf | A_f(\thetabf_{rd})-A_i(\thetabf_{insp}) < 0)$, which was found to be $ 1.21\times 10^{-6} $. There were a total of 3129 importance samples satisfying this criterion. To test if $\hat{p}_{H_1}$ is reliably estimated, I compute both the variance of $\hat{p}_{H_1}$ and the effective equal-weight sample size following the discussions in \cite{prasad_spins}. I find the relative error $E_{\hat{p}} = std(\hat{p}_{H_1})/\hat{p}_{H_1}$ to be $\approx 0.7$, indicating that the posterior mass in this region $H_1: \Delta A < 0$ is only accurate to about one standard deviation. Converting this p-value to significance, I find that $H_0$ can be rejected at $4.61 ^{+0.24} _{-0.11}\sigma$. 

A measure of the resolution of the area increase in the observation can be obtained by computing the ratio of the mean of the prediction for the area increase from the consistency test to its standard deviation. This turns out to be $\approx 5.42$ for GW250114 in this setup.
\begin{figure}
    \centering
    \includegraphics[width=\linewidth]{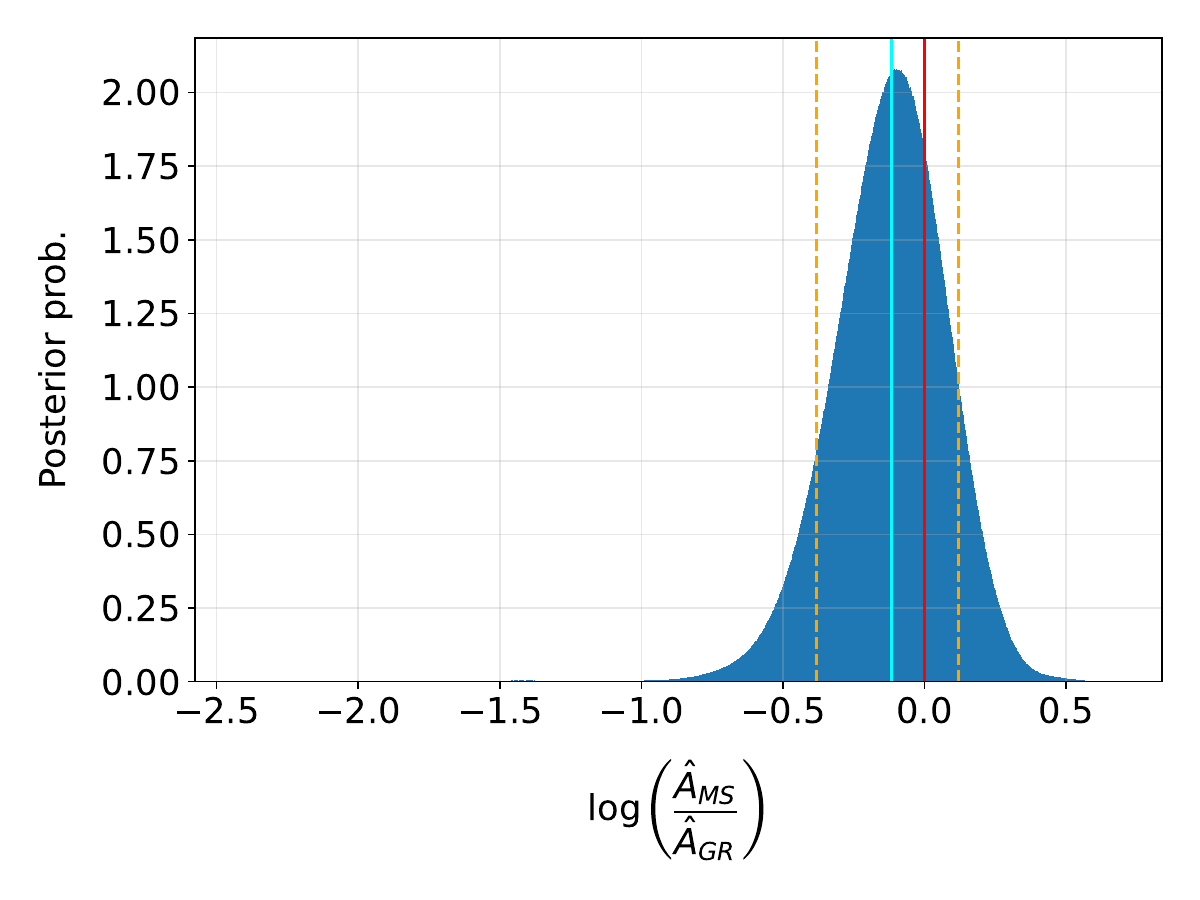}
    \caption{The ratio of the logarithms of the area increase predicted by the MSCT to that from full IMR analysis. The orange-dotted lines indicate the 10\% and 90\% quantiles. The median is indicated in cyan, and the GR value, i.e., 0, is represented by the red solid line.}
    \label{fig:ainc_test}
\end{figure}
A highly statistically significant result for area increase is a weak test of GR, and can be obeyed by many alternative theories of gravity, and can easily be affected by start time, and mode content choices, as discussed in \cite{prasad_area}. As proposed there, I instead compare the quantitative increase to the GR prediction to test GR, which is a strong test. The predictions for the area increase from GR are estimated by analyzing the entire time-domain signal without deleting any part of it and are shown in Fig.~\ref{fig:ainc}. The areas obtained in the two-segment consistency test are compared with the GR prediction in Fig.~\ref{fig:ainc_test}. Given that the posterior mass in the region $\Delta A$ is very small, for convenience, I use a log-ratio statistic $\hat{X} = \log {\dfrac{\Delta \hat{A}_{R-IM}}{\Delta \hat{A}_{GR}}}$ of the estimated area increase from the MSCT and the GR prediction from the IMR analysis to test the consistency of the observed area increase with GR. This has various advantages, such as unboundedness at both ends, negative values corresponding to the reciprocal of the argument of the log, a reduction to fractional deviation for small deviations, and greater sensitivity in the tails. The distribution has 90\% bounds at $-0.12_{-0.35}^{+0.3}$ about the median, a standard deviation of 0.21, and a GR value that lies $0.56\sigma$ away from the median, indicating that the area increase is fully consistent with GR.

\begin{figure}
    \centering
    \includegraphics[width=\linewidth]{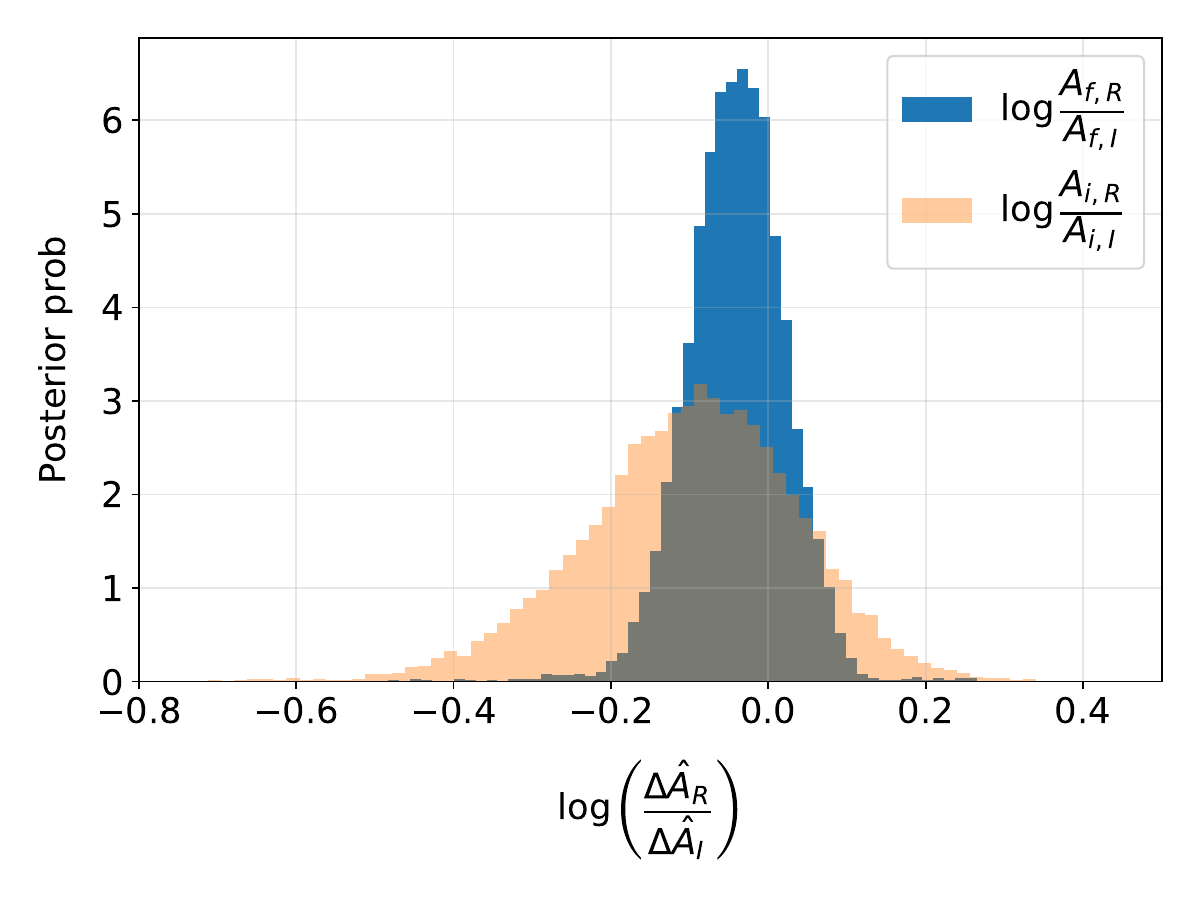}
    \caption{Consistency of the (semi-)independent predictions for the initial and final areas from each segment. The GR value is well within the 90\% confidence interval.}
    \label{fig:actseg}
\end{figure}
As a by-product of this procedure, one can also ask whether the initial areas predicted by the post-merger waveform are consistent with the initial areas computed by the inspiral analysis, and vice versa. These are shown in Fig.~\ref{fig:actseg}. The prediction of the final areas is more constrained than that of the initial areas, as intuition would suggest. In both cases, the GR value (0) fell within the 90\% CI.
\begin{figure*}
    \centering
    \includegraphics[width=0.49\linewidth]{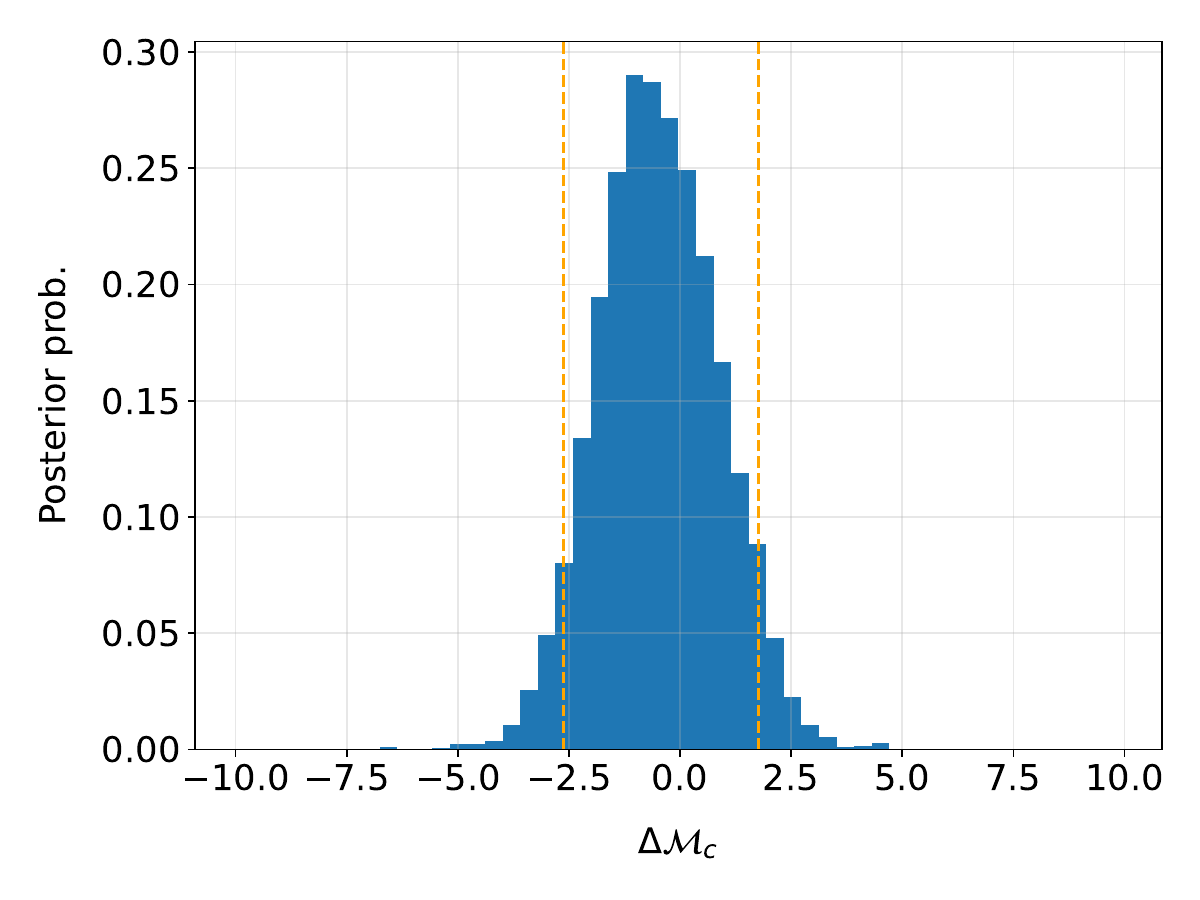}
    \includegraphics[width=0.49\linewidth]{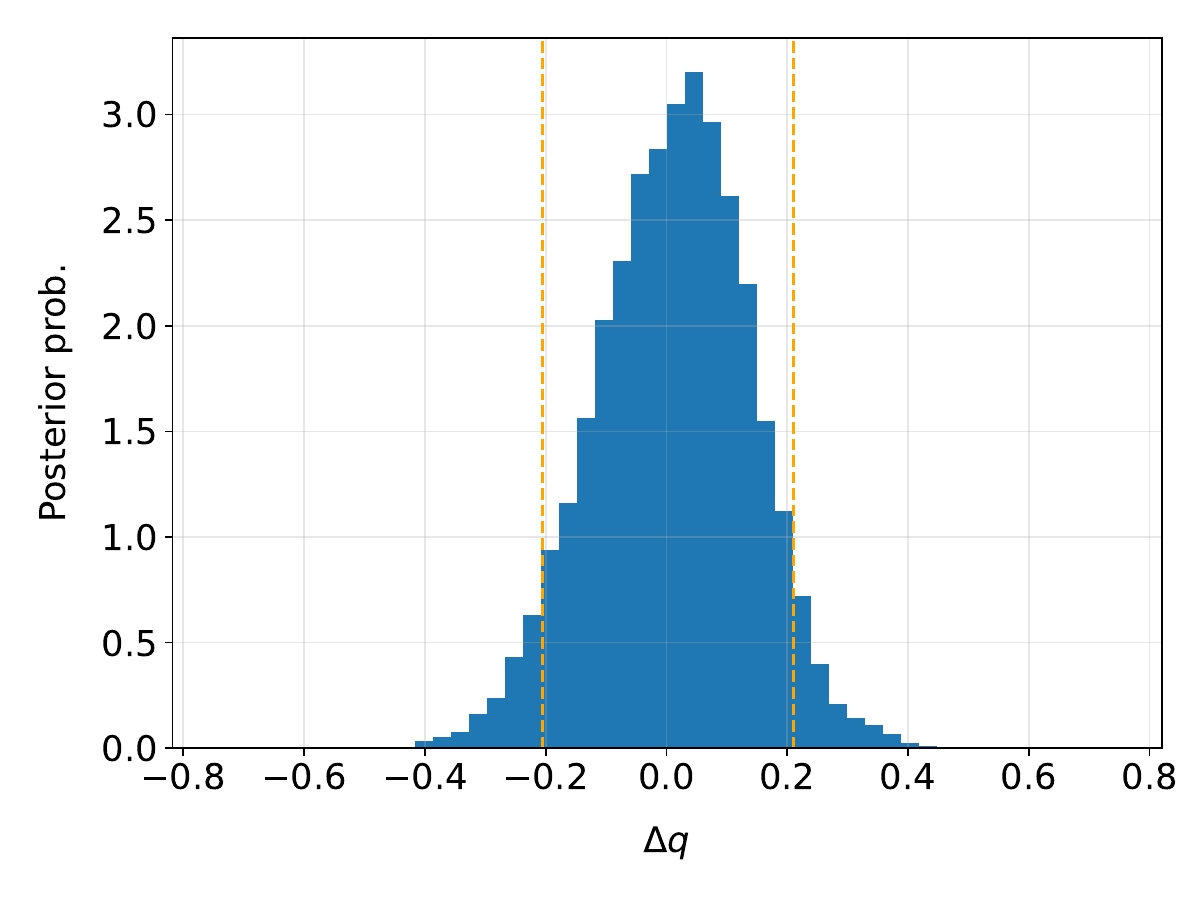}\\
    \includegraphics[width=0.49\linewidth]{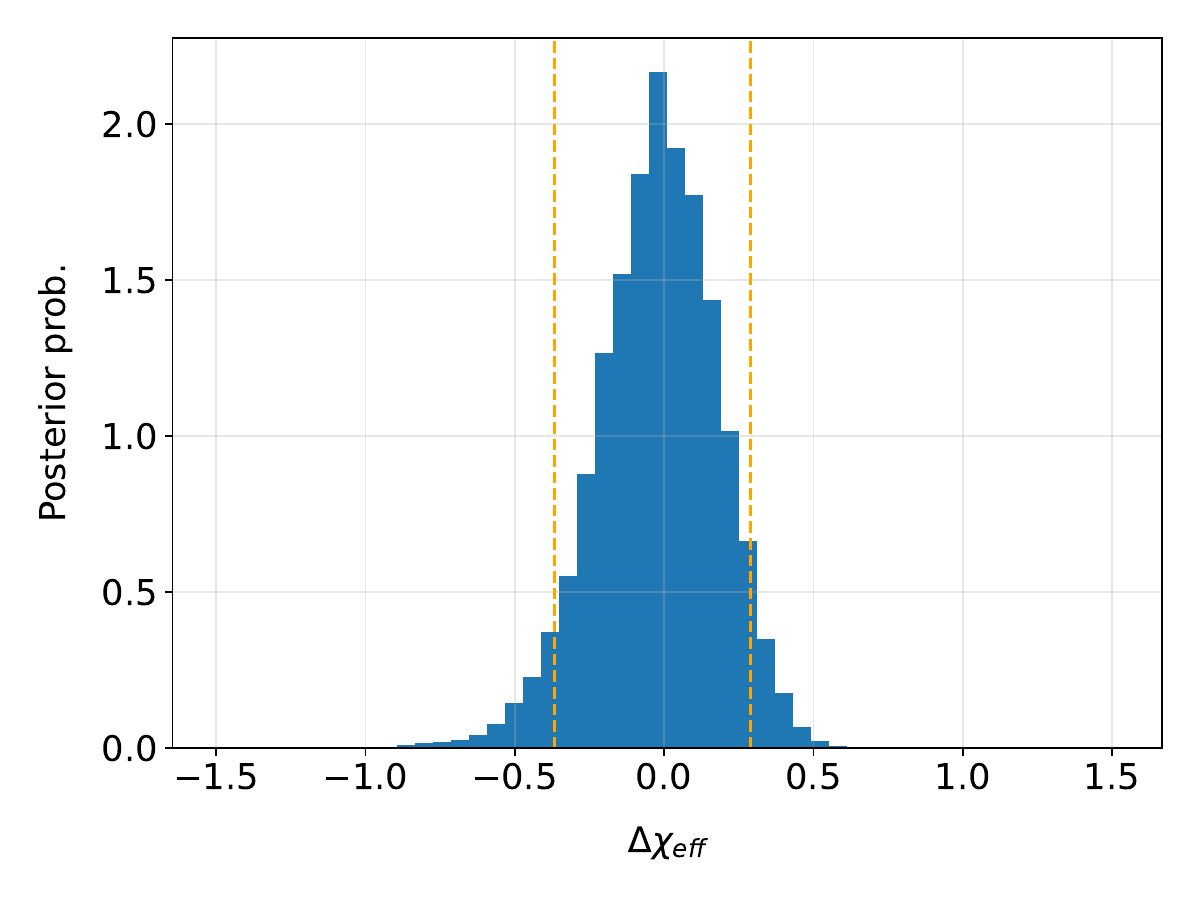}
    \includegraphics[width=0.49\linewidth]{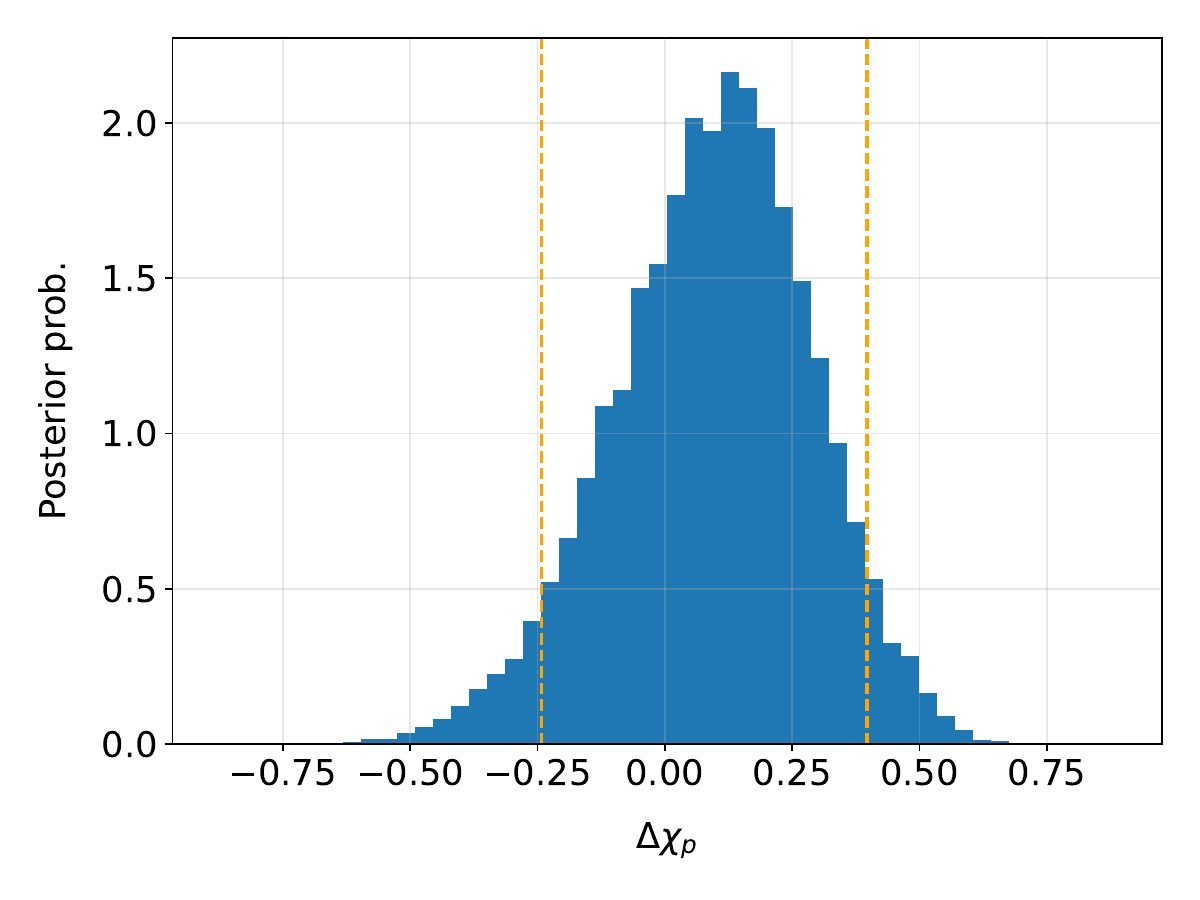}
    \caption{The consistency of the marginal distributions of four of the 10 exclusive parameters across the segments. The bounds are presented in Tab.~\ref{tab:diff_pars_and_pars}}
    \label{fig:delta}
\end{figure*}
\begin{figure*}
    \centering
    \includegraphics[width=0.49\linewidth]{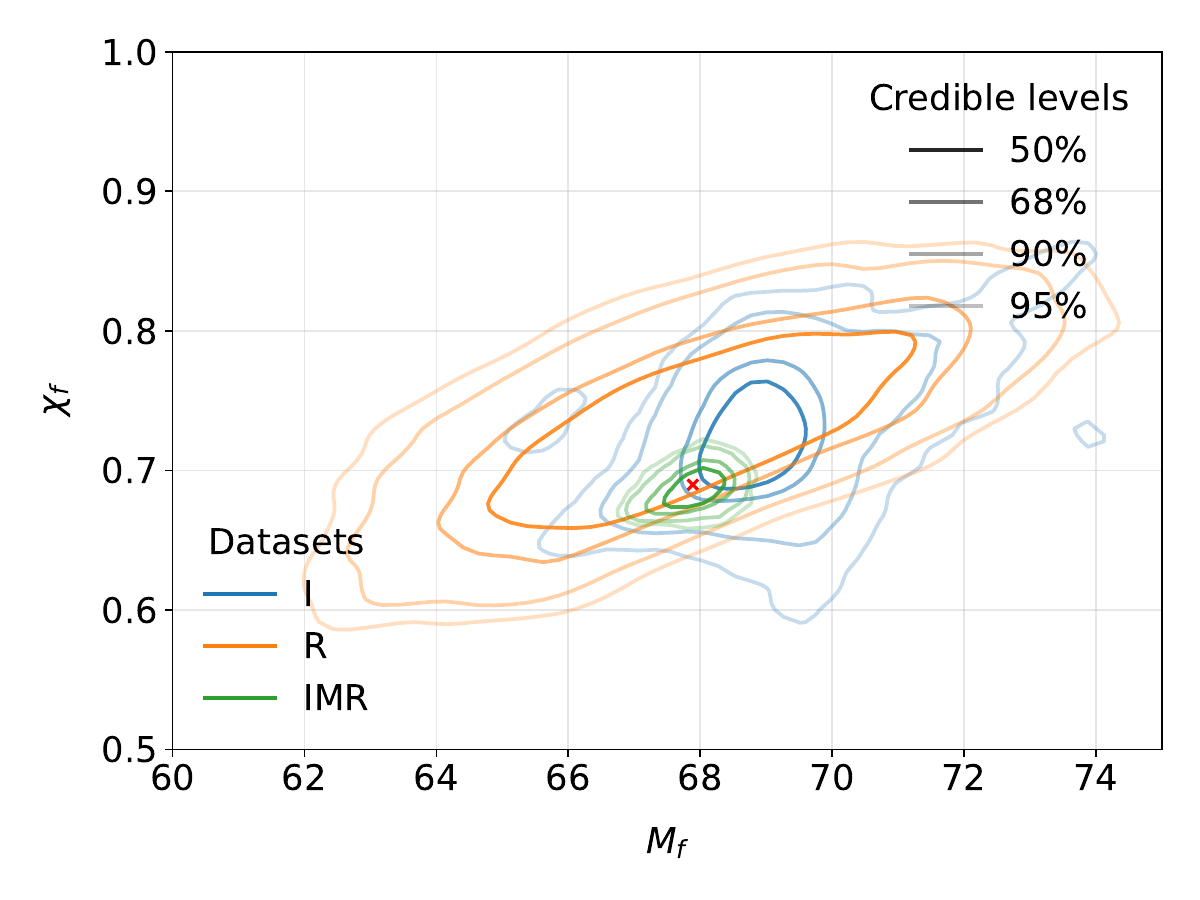}
    \includegraphics[width=0.49\linewidth]{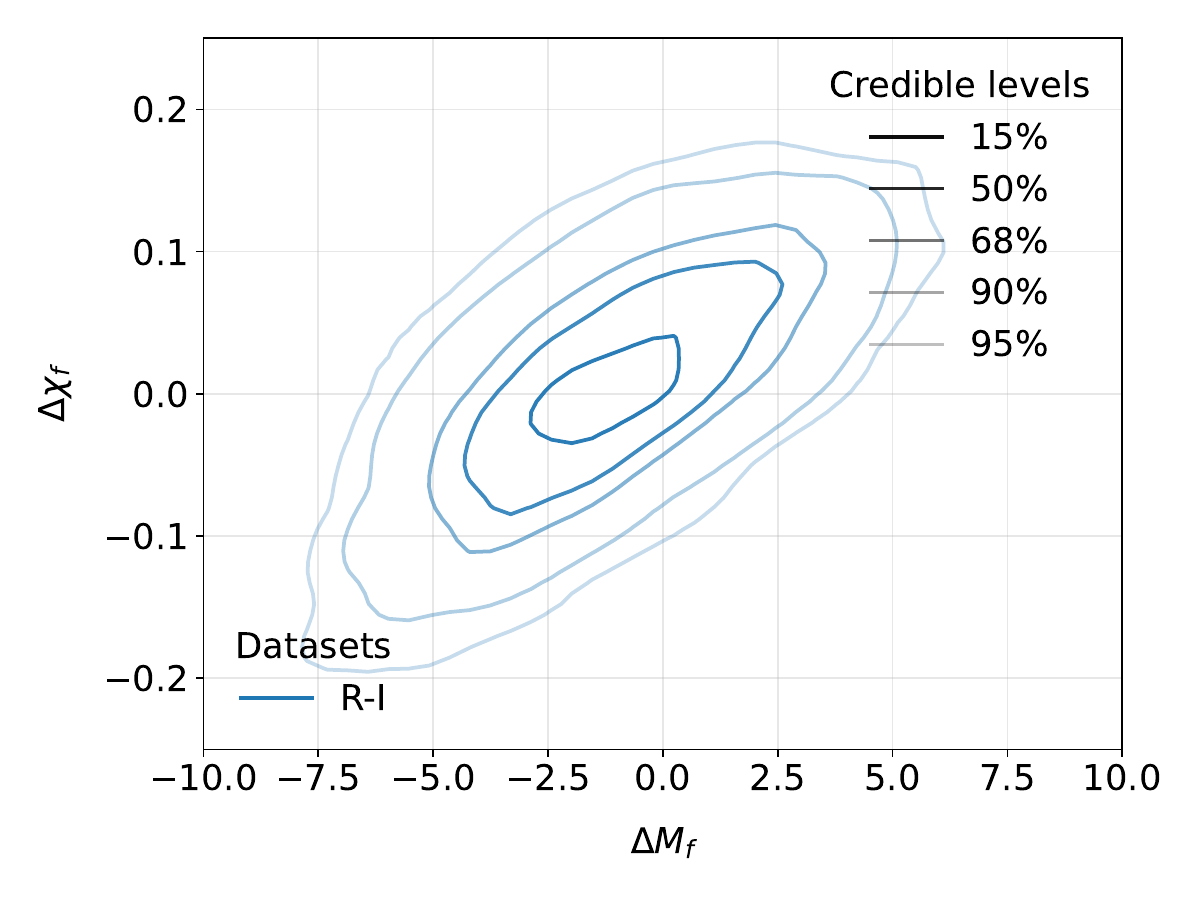}
    \caption{Consistency in the ringdown parameters. The left panel shows the 50, 68, 90, and 95\% configdence contours for the predicted final mass and spin in the I, R, and IMR analysis modes. The red cross indicates the maximum-likelihood waveform from the full IMR time-domain analysis. The right panel shows the difference between the predictions from the two segments. The GR value 0 lies at at boundary of the $15\%$ CI regions, which corresponds to $\approx 0.2 \sigma$}
    \label{fig:rd}
\end{figure*}
The posterior distributions of the 10-dimensional difference parameters contain much more information than the projected single-parameter area estimator. The marginal distributions for some are shown in Fig.~\ref{fig:delta}. 

Through this method, one can study the effects of non-modeling of physical features such as eccentricity, precession, certain post-Newtonian effects, and higher-order modes. If one analyzes the signal using waveforms that do not model these effects, the consistency between the inspiral and ringdown analyses would break down, rendering the statistical properties of the posterior distribution of the discrepancy between the segment parameters inconsistent with zero. Such a discrepancy, if it surfaces, along with the assumption that GR is valid, can be used to \emph{detect} such effects in the signal robustly.

Further, in the same spirit as the IMRCT, it is possible to study the consistency between the final mass and final spin parameters by choosing the $\hat{X}$ correspondingly. In ~\ref{fig:rd}, I show the contours for various confidence intervals for the predicted final mass and spin of GW250114, from each segment as well as the full IMR analysis. In the right panel, I also show the contours of the difference between the inspiral and ringdown predictions.

While BH spectroscopy has its own advantages and tests other aspects of the BBH ringdown waveform, a key advantage of formulating the consistency test without using BH spectroscopy is that the physical correctness of the test statistic (area increase) is independent of the choice of the start time of the ringdown analysis. In this approach, by fixing the duration and start time of the signal segments and sampling over the peak time, which is chosen as a common parameter, I allow the waveform model to dynamically determine the portion of the ringdown signal based on the likelihood values. This approach finds the best-fitting waveform in each segment and optimizes over the signal's peak time. Furthermore, the area-increase statistic is sensitive to the number of quasi-normal modes included in the analysis, as shown in \cite{prasad_area}. This method mitigates the issue by using waveform models that include all quasi-normal modes, their overtones, and the nonlinear modes currently modeled in the chosen waveform model, regardless of the signal's linear ringdown start time. Thus, the accuracy of the analysis in each segment is determined by the accuracy of the waveform models in that segment, simplifying the error budgets.

\paragraph{Computing Cost}

All production runs were conducted on the PSC Bridges2 HPC \cite{Brown2021Bridges2}, with the developmental work carried out also on the NCSA Delta system~\cite{DeltaNCSA}. The computational cost of each IMRCT parameter estimation run, including post-processing and area test costs, is approximately 15,000 CPU hours. This equates to 5 days of wall time on 128 cores.

The traditional IMRCT costs about 4000 CPU hours at 2000 live points and dlogz~$0.01$. The approximately $7.5\times$ increase in cost is expected due to the comprehensive nature of the MSCT: increase in dimensionality $D$ from 15 to 25 dimensions, which usually scales as $D^{1.5\sim2}\approx2.6\times$ without multi-modalities. The \emph{tdanalysis} accelerated time-domain pipeline is $2\times$ slower than traditional frequency-domain methods. 

This theoretically totals to $\approx (n^2 + n)\times$ computational cost for an $n$-segment MSCT, which is about $5\times$ the computational cost when $n=2$, consistent with what is observed.

This increase in cost is within reach for population analysis as well, given the increasing capabilities of computational infrastructure. 

\paragraph{Stability}

The accelerated time-domain pipeline developed in \emph{tdanalysis}~\cite{prasad_tda} is tested for computational stability under perturbations in the locations of the segment boundaries. Small perturbations $~5M$ lead to almost the same posterior distributions. This is also found to carry over to MSCT.

I perform several injections, both in zero noise and in simulated noise, to study the recovery behaviour of the 2-segment test. All these studies are consistent with expectations, and no significant systematic biases are observed. In the end matter of Sec.~\ref{sec:inj}, I discuss one such injection result for an extreme scenario in which the ringdown has an SNR of $\sim7$.

\paragraph{Conclusions}

With more and more events being detected at high SNRs, testing limited portions of the gravitational-wave signal is now feasible and an interesting endeavor. This would allow us not only to test whether individual segments of the signal agree with General Relativity, but also to assess whether the analysis is consistent across the segments. Time-domain analysis is well-suited to this endeavor. The time-domain segment-consistency likelihood model and test presented here acknowledge that the different parts of the signal being analyzed for consistency come from the same source and produce an n-dimensional consistency test, where n is the number of exclusive parameters chosen for the individual segments. The area law test of General Relativity can be viewed as a one-dimensional projection of this multi-dimensional, multi-segment consistency test of General Relativity, and provides a robust, convenient way to test future gravitational-wave signals. By accounting for source consistency, I find that GW250114 imposes stricter constraints on the area increase, allowing us to test dynamical nonlinear GR with unprecedented precision.

\begin{acknowledgments}
This document has been assigned the pre-print number LIGO-P2600015.

The author thanks B.S. Sathyaprakash for continuous support and encouragement.
The author also thanks Nathan Johnson-McDaniel, Swetha Bhagwat, and Koustav Chandra for feedback on an early version of this manuscript. 

This research is supported by National Science Foundation grants Nos.\, PHY-2308886, and No. PHY-2309064, and PHY-250294. This material is based upon work supported by NSF's LIGO Laboratory, which is a major facility fully funded by the National Science Foundation

Facilities: EGO:Virgo, GEO600, Kamioka:KAGRA, LIGO

Software: Plots were prepared with Matplotlib \cite{Hunter2007Matplotlib}, waveforms were generated through LALSuite \cite{LVK2018LALSuite, Wette2020SWIGLAL}.  NumPy \cite{harris2020array}, SciPy \cite{Virtanen2020SciPy} were used for data processing in generating the figures and quantities in the manuscript. Bilby \cite{Ashton2019Bilby, RomeroShaw2020BilbyValidation, Smith2020pBilby} and Dynesty \cite{Speagle2020dynesty, Koposov2025dynesty_v3_0_0} for stochastic sampling with Bayesian Inference. The time domain code used was developed by the author, beginning with his PhD thesis.

Computing: The authors acknowledge the computational resources provided by the NSF Access grant PHY-250294, and partly by the LIGO Laboratory’s CIT cluster, which is supported by National Science Foundation Grants PHY-0757058 and PHY-0823459, and the Gwave cluster at Pennsylvania State University (supported by NSF grants OAC-2346596, OAC-2201445, OAC-2103662, OAC-2018299, and PHY-2110594).

This work primarily used Bridges-2 at the Pittsburgh Supercomputing Center through allocation PHY-250294 from the Advanced Cyberinfrastructure Coordination Ecosystem: Services \& Support (ACCESS) program, which is supported by the National Science Foundation grants No. 2138259, 2138286, 2138307, 2137603, and 2138296. This research also used the Delta advanced computing and data resource, which is supported by the National Science Foundation (award OAC 2005572) and the State of Illinois. Delta is a joint effort of the University of Illinois Urbana-Champaign and its National Center for Supercomputing Applications.

\end{acknowledgments}

\section{End Matter}

\paragraph{Parameter Estimation settings}

All the runs used the dynesty nested sampler backend with bilby acceptance-walk for proposals, with `multi` bound method. The stopping criterion was $dlogz=0.01$. Uniform in volume priors were used for spins.




\begin{table}[t]
\centering
\renewcommand{\arraystretch}{1.5}
\begin{tabular}{l c}
\toprule
Parameter & Median ($^{+\,\mathrm{upper}}_{-\,\mathrm{lower}}$) \\
\midrule
\multicolumn{2}{l}{\textbf{Differences}}\\
$\Delta \chi_{\mathrm{eff}}$ & $-1.37\times10^{-2}{}^{+0.30}_{-0.35}$ \\
$\Delta \chi_{\mathrm{p}}$   & $0.11{}^{+0.29}_{-0.35}$ \\
$\Delta \mathcal{M}$         & $-0.52{}^{+2.29}_{-2.10}M_\odot$ \\
$\Delta q$                   & $1.68\times10^{-2}{}^{+0.19}_{-0.22}$ \\
$\Delta \phi$                & $-1.12{}^{+4.52}_{-3.02}$ \\
$\Delta \psi$                & $-0.75{}^{+1.51}_{-1.06}$ \\
$\Delta \phi_{12}$           & $-0.67{}^{+3.23}_{-2.81}$ \\
$\Delta \phi_{JL}$           & $-0.77{}^{+2.77}_{-2.96}$ \\
$\Delta \theta_{1}$          & $6.16\times10^{-2}{}^{+0.92}_{-0.88}$ \\
$\Delta \theta_{2}$          & $0.12{}^{+0.80}_{-0.80}$ \\
$\Delta M_f$                 & $-1.17{}^{+4.51}_{-4.49}M_\odot$  \\
$\Delta \chi_f$              & $7.84\times10^{-3}{}^{+0.11}_{-0.12}$ \\
\midrule
\multicolumn{2}{l}{\textbf{Parameters}}\\
$d_L$         & $5.11\times10^{2}{}^{+58.23}_{-81.15}$ \\
$\theta_{JN}$ & $2.71{}^{+0.27}_{-0.29}$ \\
$t_c$         & $1420878141.2360435 s^{+1.8ms}_{-1.7ms}$ \\
$\alpha$      & $5.45{}^{+0.19}_{-0.25}$ \\
$\delta$      & $-0.42{}^{+0.22}_{-0.14}$ \\
\bottomrule
\end{tabular}
\caption{Medians and the 10, 90\% quantiles on the marginal distribution for the deviation parameters from across the segments, and those of the common parameters are summarized here.}
\label{tab:diff_pars_and_pars}
\end{table}

\bibliography{references_norm}

@Article{Teukolsky:1972my,
    author = "Teukolsky, S. A.",
    title = "{Rotating black holes - separable wave equations for gravitational and electromagnetic perturbations}",
    journal = "Phys. Rev. Lett.",
    volume = "29",
    year = "1972",
    pages = "1114-1118",
    doi = "10.1103/PhysRevLett.29.1114",
    SLACcitation = "\\%\\%CITATION = PRLTA,29,1114;\\%\\%"
}

@Article{Dhurandhar:1992mw,
    author = "Dhurandhar, S.V. and Sathyaprakash, B.S.",
    title = "{Choice of filters for the detection of gravitational waves from coalescing binaries. 2. Detection in colored noise}",
    journal = "Phys. Rev. D",
    volume = "49",
    year = "1994",
    pages = "1707-1722",
    doi = "10.1103/PhysRevD.49.1707",
    SLACcitation = "\\%\\%CITATION = PHRVA,D49,1707;\\%\\%"
}

@Article{Sathyaprakash:1991mt,
    author = "Sathyaprakash, B.S. and Dhurandhar, S.V.",
    title = "{Choice of filters for the detection of gravitational waves from coalescing binaries}",
    journal = "Phys. Rev. D",
    volume = "44",
    year = "1991",
    pages = "3819-3834",
    doi = "10.1103/PhysRevD.44.3819",
    SLACcitation = "\\%\\%CITATION = PHRVA,D44,3819;\\%\\%"
}

@article{Living:Blanchet,
    author = "Blanchet, Luc",
    title = "Gravitational Radiation from Post-Newtonian Sources and Inspiralling Compact Binaries",
    journal = "Living Reviews in Relativity",
    year = "2006",
    number = "4",
    volume = "9",
    doi = "10.12942/lrr-2006-4",
    url = "(Cited on 08 October 2013) http://www.livingreviews.org/lrr-2006-4"
}

@Article{Cutler:1994ys,
    author = "Cutler, C. and Flanagan, \'E.\'E.",
    title = "{Gravitational waves from merging compact binaries: How accurately can one extract the binary's parameters from the inspiral wave form?}",
    journal = "Phys. Rev.",
    volume = "D49",
    year = "1994",
    pages = "2658-2697",
    eprint = "gr-qc/9402014",
    archivePrefix = "arXiv",
    doi = "10.1103/PhysRevD.49.2658",
    SLACcitation = "\\%\\%CITATION = GR-QC/9402014;\\%\\%"
}

@Article{Finn:1992xs,
    author = "Finn, L.S. and Chernoff, D.F.",
    title = "Observing binary inspiral in gravitational radiation: One interferometer",
    journal = "Phys. Rev. D",
    volume = "47",
    year = "1993",
    pages = "2198-2219",
    eprint = "gr-qc/9301003",
    SLACcitation = "\\%\\%CITATION = GR-QC/9301003;\\%\\%"
}

@Article{Finn:1992wt,
    author = "Finn, L.S.",
    title = "Detection, measurement and gravitational radiation",
    journal = "Phys. Rev. D",
    volume = "46",
    year = "1992",
    pages = "5236-5249",
    eprint = "gr-qc/9209010",
    SLACcitation = "\\%\\%CITATION = GR-QC/9209010;\\%\\%"
}

@article{GW150914,
    author = "Abbott, B. P. and others",
    title = "Observation of Gravitational Waves from a Binary Black Hole Merger",
    collaboration = "LIGO Scientific Collaboration and Virgo Collaboration",
    journal = "Phys. Rev. Lett.",
    volume = "116",
    issue = "6",
    pages = "061102",
    numpages = "16",
    year = "2016",
    month = "Feb",
    publisher = "American Physical Society",
    doi = "10.1103/PhysRevLett.116.061102",
    url = "http://link.aps.org/doi/10.1103/PhysRevLett.116.061102"
}

@unpublished{LIGOScientific:2025hdt,
    author = "Abac, A. G. and others",
    collaboration = "LIGO Scientific, VIRGO, KAGRA",
    title = "{GWTC-4.0: An Introduction to Version 4.0 of the Gravitational-Wave Transient Catalog}",
    eprint = "2508.18080",
    archivePrefix = "arXiv",
    primaryClass = "gr-qc",
    reportNumber = "LIGO-P2400293",
    month = "8",
    note = "Submitted for publication to ApJ Lett.",
    year = "2025"
}

@article{Berti:2009kk,
    author = "Berti, Emanuele and Cardoso, Vitor and Starinets, Andrei O.",
    title = "{Quasinormal modes of black holes and black branes}",
    journal = "Class. Quant. Grav.",
    volume = "26",
    pages = "163001",
    year = "2009",
    doi = "10.1088/0264-9381/26/16/163001",
    eprint = "0905.2975",
    archivePrefix = "arXiv",
    primaryClass = "hep-th"
}

@article{Hunter2007Matplotlib,
    author = "Hunter, John D.",
    title = "Matplotlib: A 2D Graphics Environment",
    journal = "Computing in Science \\\& Engineering",
    year = "2007",
    volume = "9",
    number = "3",
    pages = "90--95",
    doi = "10.1109/MCSE.2007.55"
}

@misc{LVK2018LALSuite,
    author = "{LIGO Scientific Collaboration} and {Virgo Collaboration} and {KAGRA Collaboration}",
    title = "{LVK Algorithm Library - LALSuite}",
    year = "2018",
    doi = "10.7935/GT1W-FZ16",
    url = "https://doi.org/10.7935/GT1W-FZ16",
    note = "Free software (GPL)"
}

@article{Wette2020SWIGLAL,
    author = "Wette, Karl",
    title = "SWIGLAL: Python and Octave interfaces to the LALSuite gravitational-wave data analysis libraries",
    journal = "SoftwareX",
    volume = "12",
    year = "2020",
    pages = "100634",
    doi = "10.1016/j.softx.2020.100634",
    url = "https://doi.org/10.1016/j.softx.2020.100634"
}

@Article{harris2020array,
    author = "Harris, Charles R. and Millman, K. Jarrod and van der Walt, St{\'{e}}fan J. and Gommers, Ralf and Virtanen, Pauli and Cournapeau, David and Wieser, Eric and Taylor, Julian and Berg, Sebastian and Smith, Nathaniel J. and Kern, Robert and Picus, Matti and Hoyer, Stephan and van Kerkwijk, Marten H. and Brett, Matthew and Haldane, Allan and del R{\'{i}}o, Jaime Fern{\'{a}}ndez and Wiebe, Mark and Peterson, Pearu and G{\'{e}}rard-Marchant, Pierre and Sheppard, Kevin and Reddy, Tyler and Weckesser, Warren and Abbasi, Hameer and Gohlke, Christoph and Oliphant, Travis E.",
    title = "Array programming with {NumPy}",
    year = "2020",
    month = "September",
    journal = "Nature",
    volume = "585",
    number = "7825",
    pages = "357--362",
    doi = "10.1038/s41586-020-2649-2",
    publisher = "Springer Science and Business Media {LLC}",
    url = "https://doi.org/10.1038/s41586-020-2649-2"
}

@article{Ashton2019Bilby,
    author = "Ashton, Gregory and others",
    title = "{BILBY}: A user-friendly Bayesian inference library for gravitational-wave astronomy",
    journal = "The Astrophysical Journal Supplement Series",
    year = "2019",
    volume = "241",
    number = "2",
    pages = "27",
    doi = "10.3847/1538-4365/ab06fc",
    eprint = "1811.02042",
    archivePrefix = "arXiv",
    primaryClass = "astro-ph.IM",
    url = "https://doi.org/10.3847/1538-4365/ab06fc"
}

@article{RomeroShaw2020BilbyValidation,
    author = "Romero-Shaw, I. M. and Talbot, C. and Biscoveanu, S. and others",
    title = "Bayesian inference for compact binary coalescences with bilby: validation and application to the first {LIGO}--{Virgo} gravitational-wave transient catalogue",
    journal = "Monthly Notices of the Royal Astronomical Society",
    year = "2020",
    volume = "499",
    number = "3",
    pages = "3295--3319",
    doi = "10.1093/mnras/staa2850",
    url = "https://doi.org/10.1093/mnras/staa2850"
}

@article{Smith2020pBilby,
    author = "Smith, Rory J. E. and Ashton, Gregory and Vajpeyi, Avi and Talbot, Colm",
    title = "Massively parallel Bayesian inference for transient gravitational-wave astronomy",
    journal = "Monthly Notices of the Royal Astronomical Society",
    year = "2020",
    volume = "498",
    number = "3",
    pages = "4492--4502",
    doi = "10.1093/mnras/staa2483",
    url = "https://doi.org/10.1093/mnras/staa2483"
}

@article{Virtanen2020SciPy,
    author = "Virtanen, Pauli and Gommers, Ralf and Oliphant, Travis E. and Haberland, Matt and Reddy, Tyler and Cournapeau, David and Burovski, Evgeni and Peterson, Pearu and Weckesser, Warren and Bright, Jonathan and {van der Walt}, St{\'e}fan J. and Brett, Matthew and Wilson, Joshua and Millman, K. Jarrod and Mayorov, Nikolay and Nelson, Andrew R. J. and Jones, Eric and Kern, Robert and Larson, Eric and Carey, C. J. and Polat, {\.I}lhan and Feng, Yu and Moore, Eric W. and VanderPlas, Jake and Laxalde, Denis and Perktold, Josef and Cimrman, Robert and Henriksen, Ian and Quintero, E. A. and Harris, Charles R. and Archibald, Anne M. and Ribeiro, Ant{\^o}nio H. and Pedregosa, Fabian and {van Mulbregt}, Paul and {SciPy 1.0 Contributors}",
    title = "{SciPy} 1.0: Fundamental Algorithms for Scientific Computing in Python",
    journal = "Nature Methods",
    year = "2020",
    volume = "17",
    number = "3",
    pages = "261--272",
    doi = "10.1038/s41592-019-0686-2",
    url = "https://doi.org/10.1038/s41592-019-0686-2"
}

@article{Speagle2020dynesty,
    author = "Speagle, Joshua S.",
    title = "dynesty: a dynamic nested sampling package for estimating Bayesian posteriors and evidences",
    journal = "Monthly Notices of the Royal Astronomical Society",
    year = "2020",
    volume = "493",
    number = "3",
    pages = "3132--3158",
    doi = "10.1093/mnras/staa278",
    eprint = "1904.02180",
    archivePrefix = "arXiv",
    primaryClass = "astro-ph.IM",
    url = "https://doi.org/10.1093/mnras/staa278"
}

@misc{Koposov2025dynesty_v3_0_0,
    author = "Koposov, Sergey and Speagle, Josh and Barbary, Kyle and others",
    title = "{joshspeagle/dynesty: v3.0.0}",
    year = "2025",
    month = "October",
    version = "v3.0.0",
    publisher = "Zenodo",
    doi = "10.5281/zenodo.17268284",
    url = "https://doi.org/10.5281/zenodo.17268284"
}

@inproceedings{Brown2021Bridges2,
    author = "Brown, S. T. and Buitrago, P. and Hanna, E. and Sanielevici, S. and Scibek, R. and Nystrom, N. A.",
    title = "{Bridges-2}: A Platform for Rapidly-Evolving and Data Intensive Research",
    booktitle = "Practice and Experience in Advanced Research Computing",
    year = "2021",
    pages = "1--4",
    publisher = "ACM",
    doi = "10.1145/3437359.3465593"
}

@misc{DeltaNCSA,
  author       = {{National Center for Supercomputing Applications}},
  title        = {{Delta High-Performance Computing System}},
  year         = {2022},
  howpublished = {\url{https://delta.ncsa.illinois.edu/}},
  note         = {Advanced Cyberinfrastructure Coordination Ecosystem: Services \& Support (ACCESS)}
}

@unpublished{Pompili:2025cdc,
    author = "Pompili, Lorenzo and Maggio, Elisa and Silva, Hector O. and Buonanno, Alessandra",
    title = "{A parametrized spin-precessing inspiral-merger-ringdown waveform model for tests of general relativity}",
    eprint = "2504.10130",
    archivePrefix = "arXiv",
    primaryClass = "gr-qc",
    month = "4",
    year = "2025",
    note = "{\url{https://arxiv.org/abs/2504.10130}}"
}

@article{Carullo:2019flw,
    author = "Carullo, Gregorio and Del Pozzo, Walter and Veitch, John",
    title = "{Observational Black Hole Spectroscopy: A time-domain multimode analysis of GW150914}",
    journal = "Phys. Rev. D",
    volume = "99",
    year = "2019",
    number = "12",
    pages = "123029",
    doi = "10.1103/PhysRevD.99.123029",
    note = "[Erratum: \href{http://doi.org/10.1103/PhysRevD.100.089903}{Phys.\ Rev.\ D {\bf{100}}, 089903(E) (2019)}]",
    eprint = "1902.07527",
    archivePrefix = "arXiv",
    primaryClass = "gr-qc",
    SLACcitation = "\\%\\%CITATION = ARXIV:1902.07527;\\%\\%"
}

@article{Gennari:2023gmx,
    author = "Gennari, Vasco and Carullo, Gregorio and Del Pozzo, Walter",
    title = "{Searching for ringdown higher modes with a numerical relativity-informed post-merger model}",
    eprint = "2312.12515",
    archivePrefix = "arXiv",
    primaryClass = "gr-qc",
    doi = "10.1140/epjc/s10052-024-12550-x",
    journal = "Eur. Phys. J. C",
    volume = "84",
    number = "3",
    pages = "233",
    year = "2024"
}

@article{Scheel:2025jct,
    author = "Scheel, Mark A. and others",
    title = "{The SXS collaboration{\textquoteright}s third catalog of binary black hole simulations}",
    eprint = "2505.13378",
    archivePrefix = "arXiv",
    primaryClass = "gr-qc",
    doi = "10.1088/1361-6382/adfd34",
    journal = "Class. Quant. Grav.",
    volume = "42",
    number = "19",
    pages = "195017",
    year = "2025"
}

@article{Abbott2016_AdvancedLIGO_FirstDiscoveries,
    author = "Abbott, B. P. and others",
    title = "GW150914: The Advanced LIGO Detectors in the Era of First Discoveries",
    collaboration = "LIGO Scientific Collaboration and Virgo Collaboration",
    journal = "Physical Review Letters",
    volume = "116",
    number = "13",
    pages = "131103",
    year = "2016",
    doi = "10.1103/PhysRevLett.116.131103"
}

@article{Martynov2016_SensitivityBeginning,
    author = "Martynov, D. V. and others",
    title = "Sensitivity of the Advanced {LIGO} Detectors at the Beginning of Gravitational Wave Astronomy",
    collaboration = "LIGO Scientific Collaboration",
    journal = "Physical Review D",
    volume = "93",
    number = "11",
    pages = "112004",
    year = "2016",
    doi = "10.1103/PhysRevD.93.112004"
}

@article{Tse2019_QuantumEnhancedALIGO,
    author = "Tse, M. and others",
    title = "Quantum-Enhanced Advanced {LIGO} Detectors in the Era of Gravitational-Wave Astronomy",
    collaboration = "LIGO Scientific Collaboration",
    journal = "Physical Review Letters",
    volume = "123",
    pages = "231107",
    year = "2019",
    doi = "10.1103/PhysRevLett.123.231107"
}

@article{Acernese2019_VirgoSqueezing,
    author = "Acernese, F. and others",
    title = "Increasing the Astrophysical Reach of the Advanced {Virgo} Detector via the Application of Squeezed Vacuum States of Light",
    collaboration = "Virgo Collaboration",
    journal = "Physical Review Letters",
    volume = "123",
    pages = "231108",
    year = "2019",
    doi = "10.1103/PhysRevLett.123.231108"
}

@article{Buikema2020_ALIGO_O3Performance,
    author = "Buikema, A. and others",
    title = "Sensitivity and Performance of the Advanced {LIGO} Detectors in the Third Observing Run",
    collaboration = "LIGO Scientific Collaboration",
    journal = "Physical Review D",
    volume = "102",
    number = "6",
    pages = "062003",
    year = "2020",
    doi = "10.1103/PhysRevD.102.062003"
}

@article{Capote2025_ALIGO_O4Performance,
    author = "Capote, E. and others",
    title = "Advanced {LIGO} detector performance in the fourth observing run",
    collaboration = "LIGO Scientific Collaboration",
    journal = "Physical Review D",
    volume = "111",
    number = "6",
    pages = "062002",
    year = "2025",
    doi = "10.1103/PhysRevD.111.062002"
}

@article{VirgoCollaboration2025_AdV_O4OpticalCharacterization,
    author = "{Virgo Collaboration}",
    title = "Optical characterization of the Advanced {Virgo} gravitational wave detector for the {O4} observing run",
    journal = "Applied Optics",
    volume = "64",
    number = "17",
    pages = "4710--4726",
    year = "2025",
    doi = "10.1364/AO.555312"
}

@article{KAGRACollaboration2019_NatAstron,
    author = "{KAGRA Collaboration}",
    title = "{KAGRA}: 2.5 generation interferometric gravitational wave detector",
    journal = "Nature Astronomy",
    volume = "3",
    pages = "35--40",
    year = "2019",
    doi = "10.1038/s41550-018-0658-y"
}

@article{Akutsu2021_KAGRA_DesignHistory,
    author = "Akutsu, T. and others",
    title = "Overview of {KAGRA}: Detector design and construction history",
    collaboration = "KAGRA Collaboration",
    journal = "Progress of Theoretical and Experimental Physics",
    volume = "2021",
    number = "5",
    pages = "05A101",
    year = "2021",
    doi = "10.1093/ptep/ptaa125"
}

@article{Cooper2023_Aplus_SensorsActuators,
    author = "Cooper, S. J. and others",
    title = "Sensors and Actuators for the Advanced {LIGO}+ Upgrade",
    journal = "Review of Scientific Instruments",
    volume = "94",
    number = "1",
    pages = "014502",
    year = "2023",
    doi = "10.1063/5.0117605"
}

@article{Messick2017_GstLAL_PromptDiscovery,
    author = "Messick, Cody and others",
    title = "Analysis framework for the prompt discovery of compact binary mergers in gravitational-wave data",
    journal = "Physical Review D",
    volume = "95",
    pages = "042001",
    year = "2017",
    doi = "10.1103/PhysRevD.95.042001"
}

@article{Nitz2018_PyCBCLive,
    author = "Nitz, Alexander H. and Dal Canton, Tito and Davis, Derek and Reyes, Steven",
    title = "Rapid detection of gravitational waves from compact binary mergers with {PyCBC Live}",
    journal = "Physical Review D",
    volume = "98",
    pages = "024050",
    year = "2018",
    doi = "10.1103/PhysRevD.98.024050"
}

@article{Davies2020_ExtendingPyCBC,
    author = "Davies, Gareth S. and others",
    title = "Extending the {PyCBC} search for gravitational waves from compact binary coalescence",
    journal = "Physical Review D",
    volume = "102",
    pages = "022004",
    year = "2020",
    doi = "10.1103/PhysRevD.102.022004"
}

@article{Davis2022_PyCBC_DataQualityStreams,
    author = "Davis, Derek and Trevor, Max and Mozzon, Simone and Nuttall, Laura K.",
    title = "Incorporating information from {LIGO} data quality streams into the {PyCBC} search for gravitational waves",
    journal = "Physical Review D",
    volume = "106",
    pages = "102006",
    year = "2022",
    doi = "10.1103/PhysRevD.106.102006"
}

@article{Tsukada2023_GstLAL_RankingStats,
    author = "Tsukada, Leo and others",
    title = "Improved ranking statistics of the {GstLAL} inspiral search for compact binary mergers",
    journal = "Physical Review D",
    volume = "108",
    pages = "043004",
    year = "2023",
    doi = "10.1103/PhysRevD.108.043004"
}

@article{Cannon2021_GstLAL_SoftwareX,
    author = "Cannon, Kipp and Caudill, Sarah and Chan, Chiwai and others",
    title = "{GstLAL}: A software framework for gravitational wave discovery",
    journal = "SoftwareX",
    volume = "14",
    pages = "100680",
    year = "2021",
    doi = "10.1016/j.softx.2021.100680"
}

@article{SingerPrice2016_BAYESTAR,
    author = "Singer, Leo P. and Price, Larry R.",
    title = "Rapid Bayesian position reconstruction for gravitational-wave transients",
    journal = "Physical Review D",
    volume = "93",
    pages = "024013",
    year = "2016",
    doi = "10.1103/PhysRevD.93.024013"
}

@article{Chaudhary2024_LowLatencyAlertProducts_O4,
    author = "Chaudhary, Sushant Sharma and Toivonen, Andrew and Waratkar, Gaurav and others",
    title = "Low-latency gravitational wave alert products and their performance at the time of the fourth {LIGO}-{Virgo}-{KAGRA} observing run",
    journal = "Proceedings of the National Academy of Sciences of the United States of America",
    volume = "121",
    number = "18",
    pages = "e2316474121",
    year = "2024",
    doi = "10.1073/pnas.2316474121"
}

@article{Drago2021_cWB_SoftwareX,
    author = "Drago, M. and Gayathri, V. and Klimenko, S. and Lazzaro, C. and Milotti, E. and Mitselmakher, G. and others",
    title = "coherent {W}ave{B}urst, a pipeline for unmodeled gravitational-wave data analysis",
    journal = "SoftwareX",
    volume = "14",
    pages = "100678",
    year = "2021",
    doi = "10.1016/j.softx.2021.100678"
}

@article{Mishra2022_cWB_ML_BBH_O3,
    author = "Mishra, T. and others",
    title = "Search for binary black hole mergers in the third observing run of Advanced {LIGO}-{Virgo} using coherent {W}ave{B}urst enhanced with machine learning",
    journal = "Physical Review D",
    volume = "105",
    pages = "083018",
    year = "2022",
    doi = "10.1103/PhysRevD.105.083018"
}

@article{Szczepanczyk2023_Burst_O3_cWB_ML,
    author = "Szczepa{\'n}czyk, M. J. and others",
    title = "Search for gravitational-wave bursts in the third Advanced {LIGO}-{Virgo} run with coherent {W}ave{B}urst enhanced by machine learning",
    journal = "Physical Review D",
    volume = "107",
    pages = "062002",
    year = "2023",
    doi = "10.1103/PhysRevD.107.062002"
}

@article{Cornish2021_BayesWave_EraOfObservations,
    author = "Cornish, Neil J. and Littenberg, Tyson B. and others",
    title = "The {B}ayes{W}ave analysis pipeline in the era of gravitational wave observations",
    journal = "Physical Review D",
    volume = "103",
    pages = "044006",
    year = "2021",
    doi = "10.1103/PhysRevD.103.044006"
}

@article{Zevin2017_GravitySpy,
    author = "Zevin, Michael and Coughlin, Scott and Bahaadini, Sara and others",
    title = "Gravity Spy: Integrating Advanced {LIGO} detector characterization, machine learning, and citizen science",
    journal = "Classical and Quantum Gravity",
    volume = "34",
    pages = "064003",
    year = "2017",
    doi = "10.1088/1361-6382/aa5cea"
}

@article{Coughlin2019_NovelGlitches_DeepTransferLearning,
    author = "Coughlin, Scott and others",
    title = "Discovering novel gravitational-wave detector glitches using machine learning",
    journal = "Physical Review D",
    volume = "99",
    pages = "082002",
    year = "2019",
    doi = "10.1103/PhysRevD.99.082002"
}

@article{Garcia-Quiros2020_IMRPhenomXHM,
    author = "Garc{\'i}a-Quir{\'o}s, C. and others",
    title = "Multimode frequency-domain model for the gravitational wave signal from nonprecessing black hole binaries",
    journal = "Physical Review D",
    volume = "102",
    pages = "064002",
    year = "2020",
    doi = "10.1103/PhysRevD.102.064002"
}

@article{Pratten2021_IMRPhenomXPHM,
    author = "Pratten, G. and others",
    title = "Computationally efficient models for the dominant and subdominant harmonic modes of precessing binary black holes",
    journal = "Physical Review D",
    volume = "103",
    pages = "104056",
    year = "2021",
    doi = "10.1103/PhysRevD.103.104056"
}

@article{Khan2020_PhenomPv3HM,
    author = "Khan, S. and others",
    title = "Including higher order multipoles in gravitational-wave models for precessing binary black holes",
    journal = "Physical Review D",
    volume = "101",
    pages = "024056",
    year = "2020",
    doi = "10.1103/PhysRevD.101.024056"
}

@article{Yu2023_IMRPhenomXODE,
    author = "Yu, H. and others",
    title = "Accurate and efficient waveform model for precessing binary black holes",
    journal = "Physical Review D",
    volume = "108",
    pages = "064059",
    year = "2023",
    doi = "10.1103/PhysRevD.108.064059"
}

@article{Ossokine2020_SEOBNRv4PHM,
    author = "Ossokine, S. and others",
    title = "Multipolar effective-one-body waveforms for precessing binary black holes: Construction and validation",
    journal = "Physical Review D",
    volume = "102",
    pages = "044055",
    year = "2020",
    doi = "10.1103/PhysRevD.102.044055"
}

@article{Ramos-Buades2023_SEOBNRv5PHM,
    author = "Ramos-Buades, A. and others",
    title = "Next generation of accurate and efficient multipolar precessing-spin effective-one-body waveforms for binary black hole coalescences",
    journal = "Physical Review D",
    volume = "108",
    pages = "124037",
    year = "2023",
    doi = "10.1103/PhysRevD.108.124037"
}

@article{Nagar2018_TEOBResumS,
    author = "Nagar, A. and others",
    title = "Time-domain effective-one-body gravitational waveforms for coalescing compact binaries with nonprecessing spins, tides and self-spin effects",
    journal = "Physical Review D",
    volume = "98",
    pages = "104052",
    year = "2018",
    doi = "10.1103/PhysRevD.98.104052"
}

@article{Akcay2021_TEOBResumSP,
    author = "Akcay, S. and others",
    title = "Hybrid post-Newtonian effective-one-body scheme for spin-precessing compact binaries",
    journal = "Physical Review D",
    volume = "103",
    pages = "024014",
    year = "2021",
    doi = "10.1103/PhysRevD.103.024014"
}

@article{Varma2019_NRSur7dq4,
    author = "Varma, V. and Field, S. E. and Scheel, M. A. and Blackman, J. and Gerosa, D. and Stein, L. C. and Kidder, L. E. and Pfeiffer, H. P.",
    title = "Surrogate models for precessing binary black hole simulations with unequal masses",
    journal = "Physical Review Research",
    volume = "1",
    pages = "033015",
    year = "2019",
    doi = "10.1103/PhysRevResearch.1.033015"
}

@article{Dietrich2019_MatterImprints,
    author = "Dietrich, T. and others",
    title = "Matter imprints in waveform models for neutron star binaries: Tidal and self-spin effects",
    journal = "Physical Review D",
    volume = "99",
    pages = "024029",
    year = "2019",
    doi = "10.1103/PhysRevD.99.024029"
}

@article{Dietrich2019_ImprovedNRTidal,
    author = "Dietrich, T. and Samajdar, A. and Khan, S. and Johnson-McDaniel, N. K. and Tichy, W. and others",
    title = "Improving the {NRTidal} model for binary neutron star systems",
    journal = "Physical Review D",
    volume = "100",
    pages = "044003",
    year = "2019",
    doi = "10.1103/PhysRevD.100.044003"
}

@article{Cao2017_SEOBNRE,
    author = "Cao, Z. and Han, W.-B.",
    title = "Waveform model for an eccentric binary black hole based on the effective-one-body-numerical-relativity formalism",
    journal = "Physical Review D",
    volume = "96",
    pages = "044028",
    year = "2017",
    doi = "10.1103/PhysRevD.96.044028"
}

@article{Chiaramello2020_EOB_Eccentric,
    author = "Chiaramello, D. and Nagar, A. and Riemenschneider, G. and others",
    title = "Faithful analytical effective-one-body waveform model for eccentric compact binary coalescences",
    journal = "Physical Review D",
    volume = "101",
    pages = "101501",
    year = "2020",
    doi = "10.1103/PhysRevD.101.101501"
}

@article{Paul2025_ESIGMAHM,
    author = "Paul, K. and others",
    title = "Eccentric, spinning, inspiral-merger-ringdown waveform model with higher modes for the detection and characterization of binary black holes",
    journal = "Physical Review D",
    volume = "111",
    pages = "084074",
    year = "2025",
    doi = "10.1103/PhysRevD.111.084074"
}

@article{Husa2016_IMRPhenomD_Base,
    author = {Husa, Sascha and Khan, Sebastian and Hannam, Mark and P{\"u}rrer, Michael and Ohme, Frank and Jim{\'e}nez Forteza, Xisco and Boh{\'e}, Alejandro},
    title = "Frequency-domain gravitational waves from nonprecessing black-hole binaries. I. New numerical waveforms and anatomy of the signal",
    journal = "Physical Review D",
    volume = "93",
    pages = "044006",
    year = "2016",
    doi = "10.1103/PhysRevD.93.044006"
}

@article{Khan2016_IMRPhenomD_Model,
    author = {Khan, Sebastian and Husa, Sascha and Hannam, Mark and P{\"u}rrer, Michael and Ohme, Frank and Jim{\'e}nez Forteza, Xisco and Boh{\'e}, Alejandro},
    title = "Frequency-domain gravitational waves from nonprecessing black-hole binaries. II. A phenomenological model for the advanced detector era",
    journal = "Physical Review D",
    volume = "93",
    pages = "044007",
    year = "2016",
    doi = "10.1103/PhysRevD.93.044007"
}

@article{Hannam2014_PhenomP,
    author = {Hannam, Mark and Schmidt, Patricia and Boh{\'e}, Alejandro and Haegel, Leila and Husa, Sascha and Ohme, Frank and Pratten, Geraint and P{\"u}rrer, Michael},
    title = "Simple Model of Complete Precessing Black-Hole-Binary Gravitational Waveforms",
    journal = "Physical Review Letters",
    volume = "113",
    pages = "151101",
    year = "2014",
    doi = "10.1103/PhysRevLett.113.151101"
}

@article{Schmidt2015_ChiP_SingleSpinPrecession,
    author = "Schmidt, Patricia and Ohme, Frank and Hannam, Mark",
    title = "Towards models of gravitational waveforms from generic binaries. II. Modelling precession effects with a single effective precession parameter",
    journal = "Physical Review D",
    volume = "91",
    pages = "024043",
    year = "2015",
    doi = "10.1103/PhysRevD.91.024043"
}

@article{Bohe2017_SEOBNRv4,
    author = "Boh{\'e}, Alejandro and others",
    title = "Improved effective-one-body model of spinning, nonprecessing binary black holes for the era of gravitational-wave astrophysics with advanced detectors",
    journal = "Physical Review D",
    volume = "95",
    pages = "044028",
    year = "2017",
    doi = "10.1103/PhysRevD.95.044028"
}

@article{Cotesta2018_SEOBNRv4HM,
    author = "Cotesta, Roberto and others",
    title = "Enriching the symphony of gravitational waves from binary black holes by tuning higher harmonics",
    journal = "Physical Review D",
    volume = "98",
    pages = "084028",
    year = "2018",
    doi = "10.1103/PhysRevD.98.084028"
}

@article{Purrer2016_SEOBNRv2_ROM,
    author = {P{\"u}rrer, Michael},
    title = "Frequency domain reduced order model of aligned-spin effective-one-body waveforms with generic mass ratios and spins",
    journal = "Physical Review D",
    volume = "93",
    pages = "064041",
    year = "2016",
    doi = "10.1103/PhysRevD.93.064041"
}

@article{Taracchini2014_SEOBNRv2,
    author = "Taracchini, Andrea and others",
    title = "Effective-one-body model for black-hole binaries with generic mass ratios and spins",
    journal = "Physical Review D",
    volume = "89",
    pages = "061502",
    year = "2014",
    doi = "10.1103/PhysRevD.89.061502"
}

@article{BuonannoDamour1999_EOB,
    author = "Buonanno, Alessandra and Damour, Thibault",
    title = "Effective one-body approach to general relativistic two-body dynamics",
    journal = "Physical Review D",
    volume = "59",
    pages = "084006",
    year = "1999",
    doi = "10.1103/PhysRevD.59.084006"
}

@article{BuonannoDamour2000_Plunge,
    author = "Buonanno, Alessandra and Damour, Thibault",
    title = "Transition from inspiral to plunge in binary black hole coalescences",
    journal = "Physical Review D",
    volume = "62",
    pages = "064015",
    year = "2000",
    doi = "10.1103/PhysRevD.62.064015"
}

@article{FlanaganHinderer2008_TidalLove,
    author = "Flanagan, {\'E}anna {\'E}. and Hinderer, Tanja",
    title = "Constraining neutron-star tidal Love numbers with gravitational-wave detectors",
    journal = "Physical Review D",
    volume = "77",
    pages = "021502",
    year = "2008",
    doi = "10.1103/PhysRevD.77.021502"
}

@article{Vines2011_Post1PN_Tides,
    author = "Vines, Justin and Flanagan, {\'E}anna {\'E}. and Hinderer, Tanja",
    title = "Post-1-Newtonian tidal effects in the gravitational waveform from binary inspirals",
    journal = "Physical Review D",
    volume = "83",
    pages = "084051",
    year = "2011",
    doi = "10.1103/PhysRevD.83.084051"
}

@article{London2018_PhenomHM,
    author = "London, Lionel and others",
    title = "First Higher-Multipole Model of Gravitational Waves from Spinning and Coalescing Black-Hole Binaries",
    journal = "Physical Review Letters",
    volume = "120",
    pages = "161102",
    year = "2018",
    doi = "10.1103/PhysRevLett.120.161102"
}

@article{Varma2019_NRHybSur3dq8,
    author = "Varma, Vijay and others",
    title = "Surrogate model of hybridized numerical relativity binary black hole waveforms",
    journal = "Physical Review D",
    volume = "99",
    pages = "064045",
    year = "2019",
    doi = "10.1103/PhysRevD.99.064045"
}

@article{Babak2017_PrecessingEOB_Validation,
    author = "Babak, Stanislav and Taracchini, Andrea and Buonanno, Alessandra",
    title = "Validating the effective-one-body model of spinning, precessing binary black holes against numerical relativity",
    journal = "Physical Review D",
    volume = "95",
    pages = "024010",
    year = "2017",
    doi = "10.1103/PhysRevD.95.024010"
}

@article{Abbott2016_TGR_GW150914,
    author = "Abbott, B. P. and others",
    title = "Tests of General Relativity with GW150914",
    collaboration = "LIGO Scientific Collaboration and Virgo Collaboration",
    journal = "Physical Review Letters",
    volume = "116",
    number = "22",
    pages = "221101",
    year = "2016",
    doi = "10.1103/PhysRevLett.116.221101"
}

@article{Abbott2021_TGR_GWTC2,
    author = "Abbott, R. and others",
    title = "Tests of general relativity with binary black holes from the second {LIGO}-{Virgo} gravitational-wave transient catalog",
    collaboration = "LIGO Scientific Collaboration and Virgo Collaboration",
    journal = "Physical Review D",
    volume = "103",
    number = "12",
    pages = "122002",
    year = "2021",
    doi = "10.1103/PhysRevD.103.122002"
}

@article{Abbott2025_TGR_GWTC3,
    author = "Abbott, R. and others",
    title = "Tests of general relativity with {GWTC}-3",
    collaboration = "LIGO Scientific Collaboration and Virgo Collaboration and KAGRA Collaboration",
    journal = "Physical Review D",
    volume = "112",
    number = "8",
    pages = "084080",
    year = "2025",
    doi = "10.1103/PhysRevD.112.084080"
}

@article{Allen2005_Chi2TF,
    author = "Allen, Bruce",
    title = "A chi-squared time-frequency discriminator for gravitational wave detection",
    journal = "Physical Review D",
    volume = "71",
    number = "6",
    pages = "062001",
    year = "2005",
    doi = "10.1103/PhysRevD.71.062001"
}

@article{YunesPretorius2009_ppE,
    author = "Yunes, Nicol{\'a}s and Pretorius, Frans",
    title = "Fundamental theoretical bias in gravitational wave astrophysics and the parameterized post-Einsteinian framework",
    journal = "Physical Review D",
    volume = "80",
    number = "12",
    pages = "122003",
    year = "2009",
    doi = "10.1103/PhysRevD.80.122003"
}

@article{Cornish2011_ppE_Bayes,
    author = "Cornish, Neil and Sampson, Laura and Yunes, Nicol{\'a}s and Pretorius, Frans",
    title = "Gravitational wave tests of general relativity with the parameterized post-Einsteinian framework",
    journal = "Physical Review D",
    volume = "84",
    number = "6",
    pages = "062003",
    year = "2011",
    doi = "10.1103/PhysRevD.84.062003"
}

@article{DelPozzo2011_TIGER,
    author = "Del Pozzo, Walter and Li, Tjonnie G. F. and Agathos, Michalis and Van Den Broeck, Chris and Vitale, Salvatore and Veitch, John and Vecchio, Alberto",
    title = "Testing general relativity using {B}ayesian model selection",
    journal = "Physical Review D",
    volume = "83",
    number = "8",
    pages = "082002",
    year = "2011",
    doi = "10.1103/PhysRevD.83.082002"
}

@article{Agathos2014_TIGERpipeline,
    author = "Agathos, Michalis and Del Pozzo, Walter and Li, Tjonnie G. F. and Van Den Broeck, Chris and Veitch, John and Vitale, Salvatore",
    title = "{TIGER}: A data analysis pipeline for testing the strong-field dynamics of general relativity with gravitational wave signals from coalescing compact binaries",
    journal = "Physical Review D",
    volume = "89",
    number = "8",
    pages = "082001",
    year = "2014",
    doi = "10.1103/PhysRevD.89.082001"
}

@article{Arun2006_PNcoeffTest,
    author = "Arun, K. G. and Iyer, B. R. and Qusailah, M. S. S. and Sathyaprakash, B. S.",
    title = "Probing the nonlinear structure of general relativity with black hole binaries",
    journal = "Physical Review D",
    volume = "74",
    number = "2",
    pages = "024006",
    year = "2006",
    doi = "10.1103/PhysRevD.74.024006"
}

@article{Ghosh2016_IMRConsistency,
    author = "Ghosh, A. and others",
    title = "Testing general relativity using golden black-hole binaries",
    journal = "Physical Review D",
    volume = "94",
    number = "2",
    pages = "021101",
    year = "2016",
    doi = "10.1103/PhysRevD.94.021101"
}

@article{Gossan2012_NoHair_Bayes,
    author = "Gossan, Sarah and Veitch, John and Sathyaprakash, B. S.",
    title = "Bayesian model selection for testing the no-hair theorem with black hole ringdowns",
    journal = "Physical Review D",
    volume = "85",
    number = "12",
    pages = "124056",
    year = "2012",
    doi = "10.1103/PhysRevD.85.124056"
}

@article{Mirshekari2012_ModDispersion,
    author = "Mirshekari, Saeed and Yunes, Nicol{\'a}s and Will, Clifford M.",
    title = "Constraining Lorentz-violating, modified dispersion relations with gravitational waves",
    journal = "Physical Review D",
    volume = "85",
    number = "2",
    pages = "024041",
    year = "2012",
    doi = "10.1103/PhysRevD.85.024041"
}

@article{Nishizawa2009_Polarizations,
    author = "Nishizawa, Atsushi and Taruya, Atsushi and Hayama, Kazuhiro and Kawamura, Seiji and Sakagami, Masa-aki",
    title = "Probing nontensorial polarizations of stochastic gravitational-wave backgrounds with ground-based laser interferometers",
    journal = "Physical Review D",
    volume = "79",
    number = "8",
    pages = "082002",
    year = "2009",
    doi = "10.1103/PhysRevD.79.082002"
}

@article{Abbott2019_GWTC1_GRTests,
    author = "Abbott, B. P. and others",
    title = "Tests of General Relativity with the Binary Black Hole Signals from the {LIGO}-{Virgo} Catalog {GWTC}-1",
    collaboration = "LIGO Scientific Collaboration and Virgo Collaboration",
    journal = "Physical Review D",
    volume = "100",
    number = "10",
    pages = "104036",
    year = "2019",
    doi = "10.1103/PhysRevD.100.104036"
}

@article{Hawking1971_AreaTheorem,
    author = "Hawking, S. W.",
    title = "Gravitational Radiation from Colliding Black Holes",
    journal = "Physical Review Letters",
    volume = "26",
    number = "21",
    pages = "1344--1346",
    year = "1971",
    doi = "10.1103/PhysRevLett.26.1344"
}

@Article{Pretorius:2005gq,
    author = "Pretorius, Frans",
    title = "{Evolution of Binary Black Hole Spacetimes}",
    journal = "Phys. Rev. Lett.",
    volume = "95",
    year = "2005",
    pages = "121101",
    eprint = "gr-qc/0507014",
    archivePrefix = "arXiv",
    doi = "10.1103/PhysRevLett.95.121101",
    SLACcitation = "\\%\\%CITATION = GR-QC/0507014;\\%\\%"
}

@article{Cabero:2017avf,
    author = "Cabero, Miriam and Capano, Collin D. and Fischer-Birnholtz, Ofek and Krishnan, Badri and Nielsen, Alex B. and Nitz, Alexander H. and Biwer, Christopher M.",
    title = "{Observational tests of the black hole area increase law}",
    journal = "Phys. Rev.",
    volume = "D97",
    year = "2018",
    number = "12",
    pages = "124069",
    doi = "10.1103/PhysRevD.97.124069",
    eprint = "1711.09073",
    archivePrefix = "arXiv",
    primaryClass = "gr-qc",
    SLACcitation = "\\%\\%CITATION = ARXIV:1711.09073;\\%\\%"
}

@article{Campanelli:2005dd,
    author = "Campanelli, Manuela and Lousto, C. O. and Marronetti, P. and Zlochower, Y.",
    title = "{Accurate evolutions of orbiting black-hole binaries without excision}",
    journal = "Phys. Rev. Lett.",
    volume = "96",
    year = "2006",
    pages = "111101",
    doi = "10.1103/PhysRevLett.96.111101",
    eprint = "gr-qc/0511048",
    archivePrefix = "arXiv",
    primaryClass = "gr-qc",
    SLACcitation = "\\%\\%CITATION = GR-QC/0511048;\\%\\%"
}

@article{Gupta:2018znn,
    author = "Gupta, Anshu and Krishnan, Badri and Nielsen, Alex and Schnetter, Erik",
    title = "{Dynamics of marginally trapped surfaces in a binary black hole merger: Growth and approach to equilibrium}",
    journal = "Phys. Rev.",
    volume = "D97",
    year = "2018",
    number = "8",
    pages = "084028",
    doi = "10.1103/PhysRevD.97.084028",
    eprint = "1801.07048",
    archivePrefix = "arXiv",
    primaryClass = "gr-qc",
    SLACcitation = "\\%\\%CITATION = ARXIV:1801.07048;\\%\\%"
}

@misc{isi2021analyzingblackholeringdowns,
    author = "Isi, Maximiliano and Farr, Will M.",
    title = "Analyzing black-hole ringdowns",
    year = "2021",
    eprint = "2107.05609",
    archivePrefix = "arXiv",
    primaryClass = "gr-qc",
    url = "https://arxiv.org/abs/2107.05609"
}

@article{LIGOScientific:2017vwq,
    author = "Abbott, B. P. and others",
    collaboration = "LIGO Scientific, Virgo",
    title = "{GW170817: Observation of Gravitational Waves from a Binary Neutron Star Inspiral}",
    eprint = "1710.05832",
    archivePrefix = "arXiv",
    primaryClass = "gr-qc",
    reportNumber = "LIGO-P170817",
    doi = "10.1103/PhysRevLett.119.161101",
    journal = "Phys. Rev. Lett.",
    volume = "119",
    number = "16",
    pages = "161101",
    year = "2017"
}

@article{Ghosh:2017gfp,
    author = "Ghosh, Abhirup and Johnson-Mcdaniel, Nathan K. and Ghosh, Archisman and Mishra, Chandra Kant and Ajith, Parameswaran and Del Pozzo, Walter and Berry, Christopher P. L. and Nielsen, Alex B. and London, Lionel",
    title = "{Testing general relativity using gravitational wave signals from the inspiral, merger and ringdown of binary black holes}",
    eprint = "1704.06784",
    archivePrefix = "arXiv",
    primaryClass = "gr-qc",
    reportNumber = "LIGO-P1700006, ICTS-2017-3",
    doi = "10.1088/1361-6382/aa972e",
    journal = "Class. Quant. Grav.",
    volume = "35",
    number = "1",
    pages = "014002",
    year = "2018"
}

@article{Narayan:2023vhm,
    author = "Narayan, Purnima and Johnson-McDaniel, Nathan K. and Gupta, Anuradha",
    title = "{Effect of ignoring eccentricity in testing general relativity with gravitational waves}",
    eprint = "2306.04068",
    archivePrefix = "arXiv",
    primaryClass = "gr-qc",
    doi = "10.1103/PhysRevD.108.064003",
    journal = "Phys. Rev. D",
    volume = "108",
    number = "6",
    pages = "064003",
    year = "2023"
}

@unpublished{Narayan:2024rat,
    author = "Narayan, Purnima and Johnson-McDaniel, Nathan K. and Gupta, Anuradha",
    title = "{Effect of Type II Strong Gravitational Lensing on Tests of General Relativity}",
    eprint = "2412.13132",
    archivePrefix = "arXiv",
    primaryClass = "gr-qc",
    month = "12",
    year = "2024",
    note = "{\url{https://arxiv.org/abs/2412.13132}}"
}

@unpublished{LIGOScientific:2025cmm,
    author = "Abac, A. G. and others",
    collaboration = "LIGO Scientific, VIRGO, Kagra",
    title = "{GW230814: investigation of a loud gravitational-wave signal observed with a single detector}",
    eprint = "2509.07348",
    archivePrefix = "arXiv",
    primaryClass = "gr-qc",
    reportNumber = "LIGO-P230814",
    month = "9",
    year = "2025",
    journal = "arXiv",
    note = "\url{https://arxiv.org/abs/2509.07348}"
}

@article{LIGOScientific:2025brd,
    author = "Abac, A. G. and others",
    collaboration = "LIGO Scientific, Virgo, KAGRA",
    title = "{GW241011 and GW241110: Exploring Binary Formation and Fundamental Physics with Asymmetric, High-spin Black Hole Coalescences}",
    eprint = "2510.26931",
    archivePrefix = "arXiv",
    primaryClass = "astro-ph.HE",
    reportNumber = "LIGO-P2500402",
    doi = "10.3847/2041-8213/ae0d54",
    journal = "Astrophys. J. Lett.",
    volume = "993",
    number = "1",
    pages = "L21",
    year = "2025"
}

@article{Vishveshwara:1970zz,
    author = "Vishveshwara, C. V.",
    title = "{Scattering of Gravitational Radiation by a Schwarzschild Black-hole}",
    doi = "10.1038/227936a0",
    journal = "Nature",
    volume = "227",
    pages = "936--938",
    year = "1970"
}

@ARTICLE{Teukolsky:1973ap,
    author = "{Teukolsky}, S. A.",
    title = "{Perturbations of a Rotating Black Hole. I. Fundamental Equations for Gravitational, Electromagnetic, and Neutrino-Field Perturbations}",
    journal = "Astrophys. J.",
    year = "1973",
    month = "October",
    volume = "185",
    pages = "635-648",
    adsurl = "http://adsabs.harvard.edu/cgi-bin/nph-bib\\_query?bibcode=1973ApJ...185..635T\\&db\\_key=AST"
}

@unpublished{LIGOScientific:2025obp,
    author = "{LIGO-Virgo-KAGRA Collaboration}",
    collaboration = "LIGO Scientific, VIRGO, KAGRA",
    title = "{Black Hole Spectroscopy and Tests of General Relativity with GW250114}",
    eprint = "2509.08099",
    archivePrefix = "arXiv",
    primaryClass = "gr-qc",
    reportNumber = "LIGO P2500461",
    month = "9",
    year = "2025",
    note = "\url{https://arxiv.org/abs/2509.08099}"
}

@article{LIGOScientific:2025rid,
    author = "Abac, A. G. and others",
    collaboration = "LIGO Scientific, Virgo, KAGRA",
    title = "{GW250114: Testing Hawking's Area Law and the Kerr Nature of Black Holes}",
    eprint = "2509.08054",
    archivePrefix = "arXiv",
    primaryClass = "gr-qc",
    reportNumber = "LIGO-P2500421",
    doi = "10.1103/kw5g-d732",
    journal = "Phys. Rev. Lett.",
    volume = "135",
    number = "11",
    pages = "111403",
    year = "2025"
}

@article{Price:1971fb,
    author = "Price, Richard H.",
    title = "{Nonspherical perturbations of relativistic gravitational collapse. 1. Scalar and gravitational perturbations}",
    doi = "10.1103/PhysRevD.5.2419",
    journal = "Phys. Rev. D",
    volume = "5",
    pages = "2419--2438",
    year = "1972"
}

@article{Colleoni:2024knd,
    author = "Colleoni, Marta and Vidal, Felip A. Ramis and Garc{\'\i}a-Quir{\'o}s, Cecilio and Ak{\c{c}}ay, Sarp and Bera, Sayantani",
    title = "{Fast frequency-domain gravitational waveforms for precessing binaries with a new twist}",
    eprint = "2412.16721",
    archivePrefix = "arXiv",
    primaryClass = "gr-qc",
    doi = "10.1103/PhysRevD.111.104019",
    journal = "Phys. Rev. D",
    volume = "111",
    number = "10",
    pages = "104019",
    year = "2025"
}

@article{Prasad:2023bwa,
    author = "Prasad, Vaishak",
    title = "{Shear at the common dynamical horizon in binary black hole mergers and its imprint in their gravitational radiation}",
    eprint = "2312.01136",
    archivePrefix = "arXiv",
    primaryClass = "gr-qc",
    doi = "10.1103/PhysRevD.111.084070",
    journal = "Phys. Rev. D",
    volume = "111",
    number = "8",
    pages = "084070",
    year = "2025"
}

@unpublished{prasad_area,
    author = "Prasad, Vaishak and B.S., Sathyaprakash and Ashtekar, Abhay",
    title = "{Testing General Relativity using Hawking's Area Law}",
    note = "in preparation",
    year = "2025"
}

@unpublished{prasad_spins,
    author = "Prasad, Vaishak and B.S., Sathyaprakash",
    title = "{Are black-hole spins truly near-zero?}",
    eprinttype = "LIGO-DCC",
    eprint = "LIGO-P2500751",
    url = "https://dcc.ligo.org/LIGO-P2500751",
    note = "Available at: https://dcc.ligo.org/LIGO-P2500751",
    year = "2026"
}

@article{Mroue:2013,
    author = "Mroue, Abdelghani H. and others",
    title = "A Catalog of 174 Binary Black Hole Simulations for Gravitational Wave Astronomy",
    journal = "Phys. Rev. Lett.",
    volume = "111",
    pages = "241104",
    year = "2013",
    doi = "10.1103/PhysRevLett.111.241104"
}

@article{Boyle:2019SXS,
    author = "Boyle, Michael and others",
    title = "The SXS Collaboration Catalog of Binary Black Hole Simulations",
    journal = "Class. Quant. Grav.",
    volume = "36",
    pages = "195006",
    year = "2019",
    doi = "10.1088/1361-6382/ab34e2"
}

@article{Jani:2016GT,
    author = "Jani, Karan and Healy, James and Clark, James A. and London, Lionel and Laguna, Pablo and Shoemaker, Deirdre",
    title = "Georgia Tech Catalog of Gravitational Waveforms",
    journal = "Class. Quant. Grav.",
    volume = "33",
    pages = "204001",
    year = "2016",
    doi = "10.1088/0264-9381/33/20/204001"
}

@article{Healy:2017RIT,
    author = "Healy, James and Lousto, Carlos O. and Zlochower, Yosef and Campanelli, Manuela",
    title = "The RIT Binary Black Hole Simulations Catalog",
    journal = "Class. Quant. Grav.",
    volume = "34",
    pages = "224001",
    year = "2017",
    doi = "10.1088/1361-6382/aa91b1"
}

@article{Healy:2019RIT2,
    author = "Healy, James and Lousto, Carlos O. and Lange, Jacob and O'Shaughnessy, Richard and Zlochower, Yosef and Campanelli, Manuela",
    title = "The Second RIT Binary Black Hole Simulations Catalog and Its Application to Gravitational Waves Parameter Estimation",
    journal = "Phys. Rev. D",
    volume = "100",
    pages = "024021",
    year = "2019",
    doi = "10.1103/PhysRevD.100.024021"
}

@article{Healy:2020RIT3,
    author = "Healy, James and Lousto, Carlos O.",
    title = "The Third RIT Binary Black Hole Simulations Catalog",
    journal = "Phys. Rev. D",
    volume = "102",
    pages = "104018",
    year = "2020",
    doi = "10.1103/PhysRevD.102.104018"
}

@article{Ferguson:2025MAYA2,
    author = "Ferguson, Deborah and others",
    title = "Second MAYA Catalog of Binary Black Hole Numerical Relativity Waveforms",
    journal = "Phys. Rev. D",
    volume = "112",
    pages = "044043",
    year = "2025",
    doi = "10.1103/gk7x-9cds"
}

@article{Dietrich:2017aum,
    author = "Dietrich, Tim and Bernuzzi, Sebastiano and Tichy, Wolfgang",
    title = "{Closed-form tidal approximants for binary neutron star gravitational waveforms constructed from high-resolution numerical relativity simulations}",
    eprint = "1706.02969",
    archivePrefix = "arXiv",
    primaryClass = "gr-qc",
    doi = "10.1103/PhysRevD.96.121501",
    journal = "Phys. Rev. D",
    volume = "96",
    number = "12",
    pages = "121501",
    year = "2017"
}

@article{Gonzalez:2023CoRe2,
    author = "Gonzalez, Alejandra and others",
    title = "Second Release of the CoRe Database of Binary Neutron Star Merger Waveforms",
    journal = "Class. Quant. Grav.",
    volume = "40",
    pages = "085011",
    year = "2023",
    doi = "10.1088/1361-6382/acc231"
}

@book{WainsteinZubakov1970_Extraction,
    author = "Wainstein, L. A. and Zubakov, V. D.",
    title = "Extraction of Signals from Noise",
    publisher = "Dover Publications",
    year = "1970"
}

@book{Helstrom1968_StatisticalTheory,
    author = "Helstrom, Carl W.",
    title = "Statistical Theory of Signal Detection",
    edition = "2",
    publisher = "Pergamon Press",
    year = "1968"
}

@article{Turin1960_MatchedFilters,
    author = "Turin, G. L.",
    title = "An Introduction to Matched Filters",
    journal = "IRE Transactions on Information Theory",
    volume = "6",
    number = "3",
    pages = "311--329",
    year = "1960",
    doi = "10.1109/TIT.1960.1057571"
}

@article{Isi2021AreaLaw,
    author = "Isi, Maximiliano and Farr, Will M. and Giesler, Matthew and Scheel, Mark A. and Teukolsky, Saul A.",
    title = "Testing the Black-Hole Area Law with Gravitational-Wave Observations",
    journal = "Physical Review Letters",
    volume = "127",
    number = "1",
    pages = "011103",
    year = "2021",
    doi = "10.1103/PhysRevLett.127.011103",
    eprint = "2012.05794",
    archivePrefix = "arXiv",
    primaryClass = "gr-qc"
}

@article{Kalman1960KalmanFilter,
    author = "Kalman, R. E.",
    title = "A New Approach to Linear Filtering and Prediction Problems",
    journal = "Transactions of the ASME--Journal of Basic Engineering",
    volume = "82",
    number = "1",
    pages = "35--45",
    year = "1960"
}

@article{Durbin1960FittingTimeSeries,
    author = "Durbin, James",
    title = "The Fitting of Time-Series Models",
    journal = "Review of the International Statistical Institute",
    volume = "28",
    number = "3",
    pages = "233--244",
    year = "1960",
    doi = "10.2307/1401322"
}

@book{BoxJenkins1970TimeSeries,
    author = "Box, George E. P. and Jenkins, Gwilym M.",
    title = "Time Series Analysis: Forecasting and Control",
    year = "1970",
    publisher = "Holden-Day"
}

@article{Maggio:2023pPMRD,
    author = "Maggio, Elisa and Silva, Hector O. and Buonanno, Alessandra and Ghosh, Abhirup",
    title = "Tests of general relativity in the nonlinear regime: A parametrized plunge-merger-ringdown gravitational waveform model",
    journal = "Physical Review D",
    volume = "108",
    pages = "024043",
    year = "2023",
    doi = "10.1103/PhysRevD.108.024043",
    eprint = "2212.09655",
    archivePrefix = "arXiv",
    primaryClass = "gr-qc"
}

@article{Forteza:2023APConsistency,
    author = "Jim{\'e}nez Forteza, Xisco and Bhagwat, Swetha and Kumar, Sumit and Pani, Paolo",
    title = "Novel Ringdown Amplitude-Phase Consistency Test",
    journal = "Physical Review Letters",
    volume = "130",
    number = "2",
    pages = "021001",
    year = "2023",
    doi = "10.1103/PhysRevLett.130.021001"
}

@article{Cadez1974_CommonAH,
    author = "Čadež, Andrej",
    title = "Apparent horizons in the two-black-hole problem",
    journal = "Annals of Physics",
    volume = "83",
    number = "2",
    pages = "449--457",
    year = "1974",
    doi = "10.1016/0003-4916(74)90206-1"
}

@article{CookAbrahams1992_HorizonStructure,
    author = "Cook, Gregory B. and Abrahams, Alan M.",
    title = "Horizon structure of initial-data sets for axisymmetric two-black-hole collisions",
    journal = "Physical Review D",
    volume = "46",
    number = "2",
    pages = "702--713",
    year = "1992",
    doi = "10.1103/PhysRevD.46.702"
}

@article{Anninos1995_HeadOnCommonAH,
    author = "Anninos, Peter and Hobill, David and Seidel, Edward and Smarr, Larry and Suen, Wai-Mo",
    title = "Head-on collision of two equal mass black holes",
    journal = "Physical Review D",
    volume = "52",
    number = "4",
    pages = "2044--2058",
    year = "1995",
    doi = "10.1103/PhysRevD.52.2044"
}

@article{CorreiaCapano:2024SkyMarginalization,
    author = "Correia, Alex and Capano, Collin D.",
    title = "Sky marginalization in black hole spectroscopy and tests of the area theorem",
    journal = "Physical Review D",
    volume = "110",
    pages = "044018",
    year = "2024",
    doi = "10.1103/PhysRevD.110.044018",
    url = "https://doi.org/10.1103/PhysRevD.110.044018",
    eprint = "2312.15146",
    archivePrefix = "arXiv",
    primaryClass = "gr-qc"
}

@article{Correia2,
    author = "Correia, Alex and Wang, Yi-Fan and Westerweck, Julian and Capano, Collin D.",
    title = "Low evidence for ringdown overtone in GW150914 when marginalizing over time and sky location uncertainty",
    journal = "Phys. Rev. D",
    volume = "110",
    issue = "4",
    pages = "L041501",
    numpages = "8",
    year = "2024",
    month = "Aug",
    publisher = "American Physical Society",
    doi = "10.1103/PhysRevD.110.L041501",
    url = "https://link.aps.org/doi/10.1103/PhysRevD.110.L041501"
}

@unpublished{Tang:2025GW230814AreaLaw,
    author = "Tang, Shao-Peng and Wang, Hai-Tian and Li, Yin-Jie and Fan, Yi-Zhong",
    title = "Verification of the Black Hole Area Law with GW230814",
    year = "2025",
    eprint = "2509.03480",
    archivePrefix = "arXiv",
    primaryClass = "gr-qc",
    note = "{\url{https://arxiv.org/abs/2509.03480}}"
}

@article{Harris1978WindowsDFT,
    author = "Harris, Fredric J.",
    title = "On the Use of Windows for Harmonic Analysis with the Discrete Fourier Transform",
    journal = "Proceedings of the IEEE",
    volume = "66",
    number = "1",
    pages = "51--83",
    year = "1978",
    doi = "10.1109/PROC.1978.10837"
}

@article{Dahlhaus1988SmallSample,
    author = "Dahlhaus, Rainer",
    title = "Small Sample Effects in Time Series Analysis: A New Asymptotic Theory and a New Estimate",
    journal = "The Annals of Statistics",
    volume = "16",
    number = "2",
    pages = "808--841",
    year = "1988",
    doi = "10.1214/aos/1176350839"
}

@article{Guillaumin2022DebiasedWhittle,
    author = "Guillaumin, Adrien and Roueff, Fran{\c c}ois and Lavancier, Fr{\'e}d{\'e}ric",
    title = "The debiased Whittle likelihood",
    journal = "Bernoulli",
    volume = "28",
    number = "3",
    pages = "1769--1797",
    year = "2022",
    doi = "10.3150/21-BEJ1406"
}

@article{Pilz2012TaperingWindowed,
    author = {Pilz, J{\"u}rgen},
    title = "Tapering of windowed time series",
    journal = "Computational Statistics \\\& Data Analysis",
    volume = "56",
    number = "12",
    pages = "3975--3986",
    year = "2012",
    doi = "10.1016/j.csda.2012.03.015"
}

@article{Wang2023DataGapsTianQin,
    author = "Wang, Hao and Mei, Jianwei and Li, En-Kun and Hu, Yan and Luo, Jun",
    title = "Dealing with data gaps in space-based gravitational-wave detection: window function method and inpainting method",
    journal = "Physical Review D",
    volume = "108",
    pages = "043007",
    year = "2023",
    doi = "10.1103/PhysRevD.108.043007",
    archivePrefix = "arXiv",
    eprint = "2303.00241",
    primaryClass = "gr-qc"
}
\section{End Matter}
\begin{figure*}[!b]
    \centering
    \includegraphics[width=0.49\linewidth]{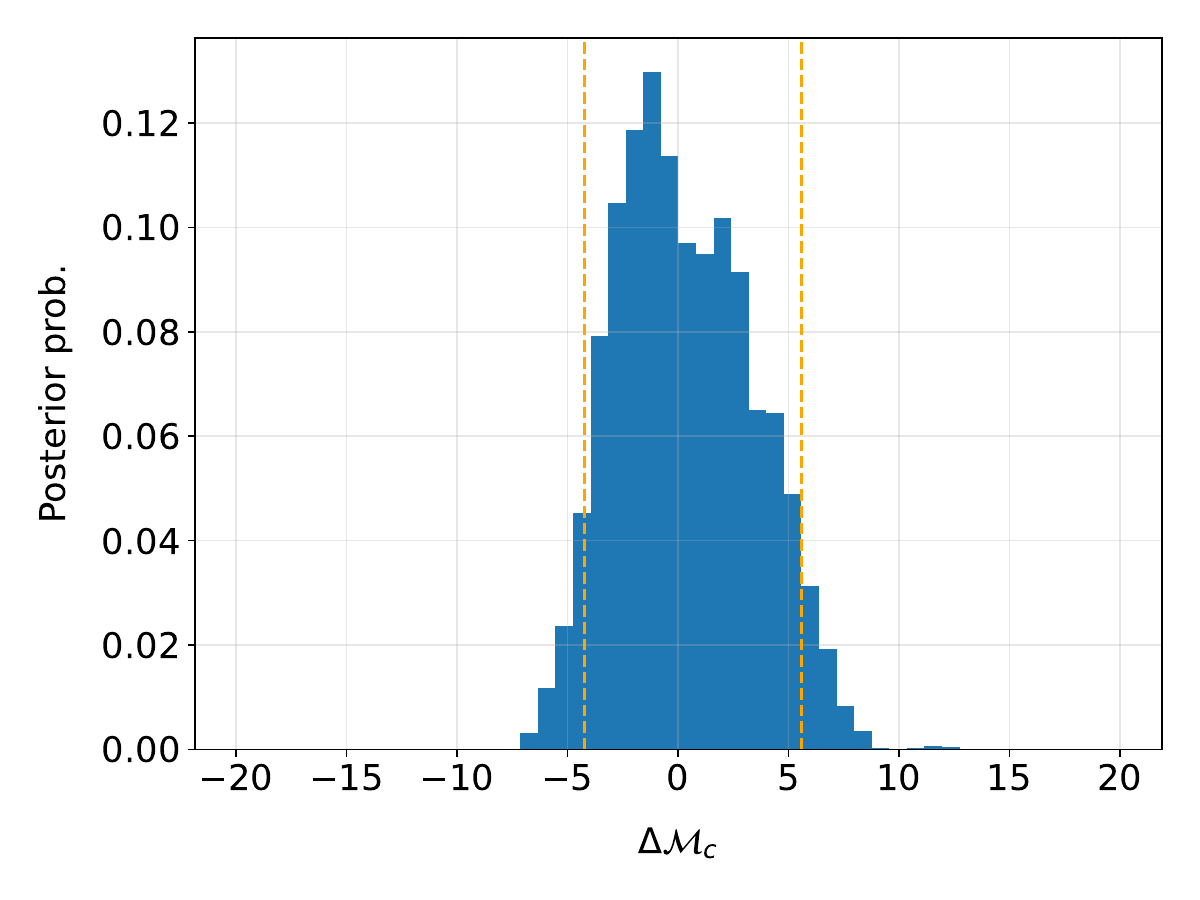}
    \includegraphics[width=0.49\linewidth]{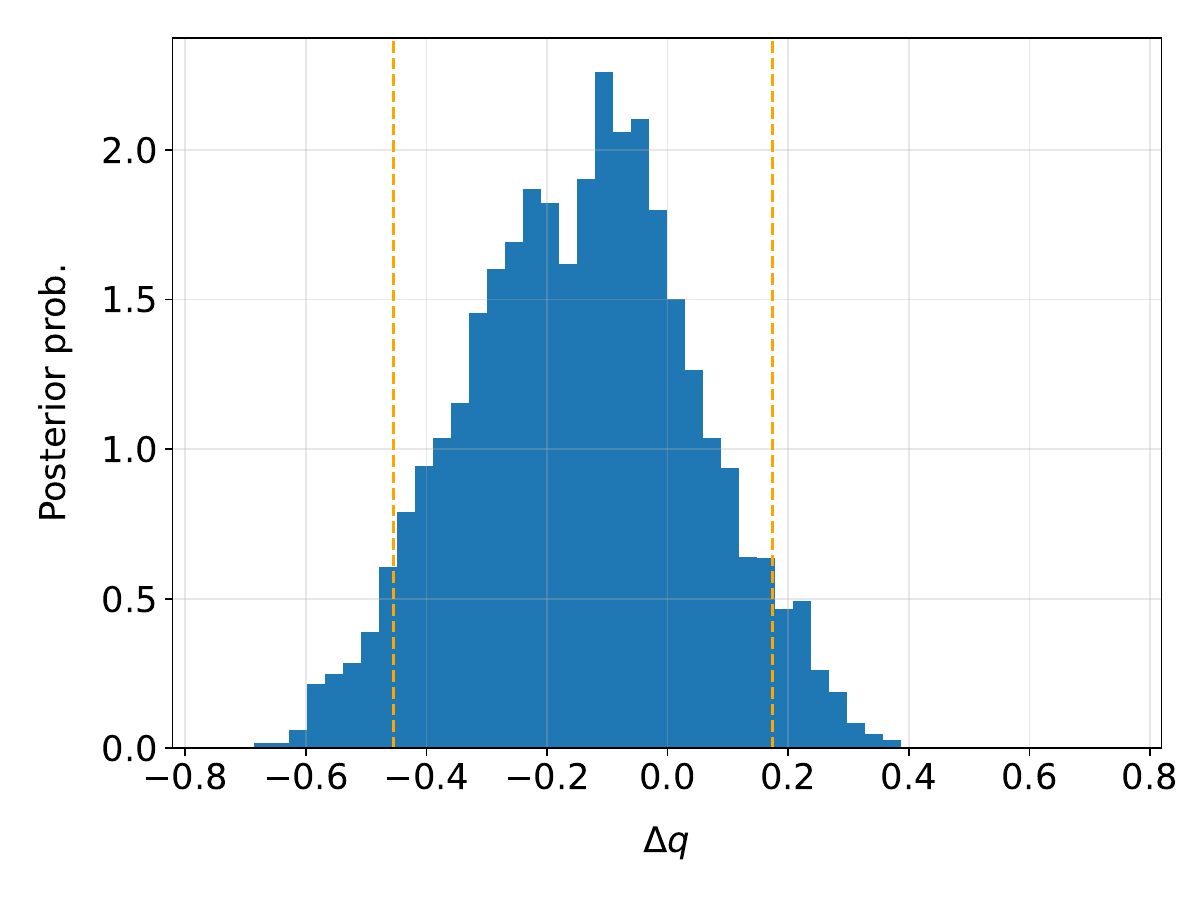}\\
    \includegraphics[width=0.49\linewidth]{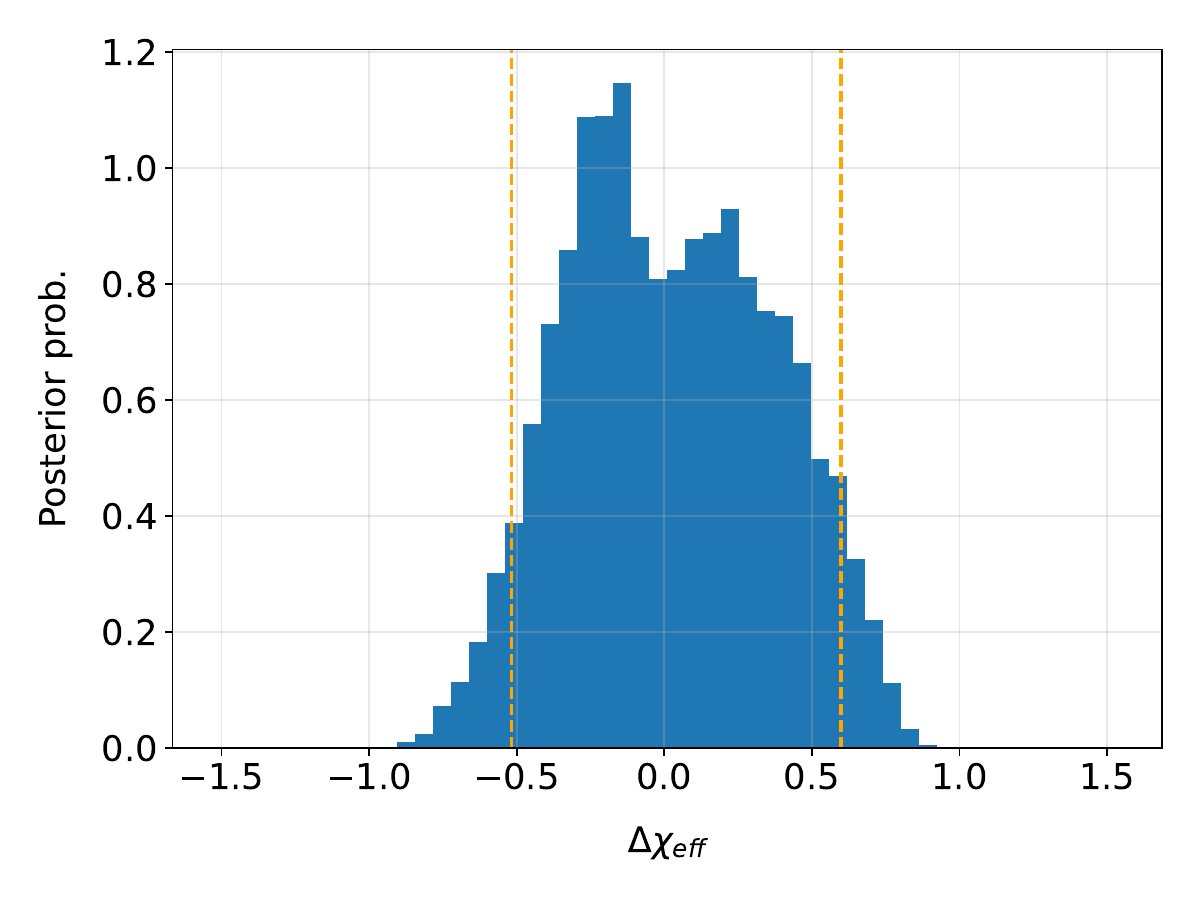}
    \includegraphics[width=0.49\linewidth]{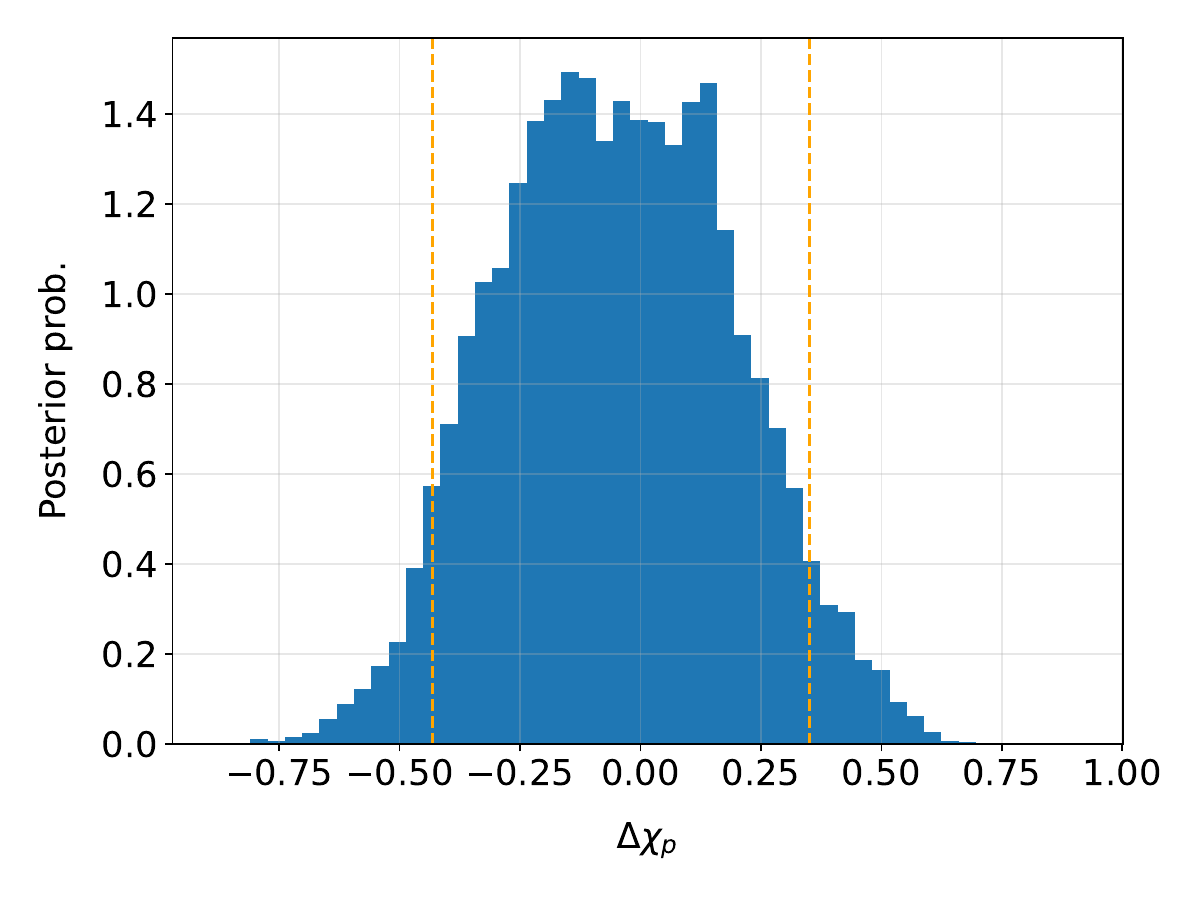}
    \caption{The 1D marginal histograms of the difference posteriors between the inspiral and ringdown analysis for the zero-noise GW250114-like signal}
    \label{fig:GW25_inj_diff}
\end{figure*}

\subsection{Singular-source constraint}
\label{sec:ss}
In this section, I discuss what singular-source means when analyzing multiple segments of the same signal, and the rationale for choosing the set $\thetabf_{c} \equiv \{\alpha, \delta, d_L, \theta_{JN}, t_{c}\}$ as the default choice rather than just $\alpha, \delta, t_{c}$. Note that the \emph{tdanalysis}~\cite{prasad_tda} framework allows any subset of the 15-dimensional parameter space to be designated as common parameters. Here, I only justify the choice made in the demonstration of the results.

The parameters $\alpha, \delta, d_L$ determine the spatial location of the source in the three-dimensional sky. For a null test of GR, it is logical to expect that the different parts of the signal are generated by the same source, and that that location would not change on the timescale of the signal being analyzed. Note that this does not mean that the overall amplitude of the signal is fixed, because the intrinsic mass parameters, including the total mass, is differently sampled across the segments.

Secondly, the peak time of the signal in the detector $t_c$ is invariant. When describing the signal using templates in General Relativity, one convenient way to represent it approximately is by the signal's polarization peak, which would not change across segments. Thus, even though waveforms with different parameters are generated by each drawn stochastic sample, keeping the value common corresponds to ensuring that the peak location of the strain in the detector is the same for each segment, while also allowing it to vary across the samples.

Thirdly, the inclination of the source in the sky denoted by the angle $\theta_{JN}$ between the line of sight and the total angular momentum vector at the reference time (or reference frequency of, say, $20Hz$) is a geometric variable that determines the orientation of the source in the sky, and the viewing angle of the observer. This is not a quantity that is specific to GR. In all theories of gravity whose weak field expansions are tensor modes, the inclination angle enters the evaluation of spin-weighted spherical harmonic modes to obtain polarizations in the same manner. If, in the theories being tested, there exist other degrees of freedom, such as vector or scalar modes, at future null infinity, then the determination of some extrinsic parameters will be biased (possibly propagating to other parameters due to correlations). However, it would still serve as a null test of GR, allowing us to answer the question: Assuming a single GR source, how consistent are the inferred GR parameters of the segments? Note that this does not imply that the degeneracies between these commonly sampled extrinsic parameters and other exclusive parameters are the same in each segment, as the exclusive parameters can be different, and they have freedom to attain values that best describe the signal in their respective segments. Tests like the IMRCT also assume GR but allow greater freedom and thus greater degeneracy in the analysis, i.e., the possibility of different source locations/orientations in the sky, which this test seeks to remedy.

Note that keeping $\theta_{JN}$ common only means that the inclination of the sources describing the inspiral and ringdown segments is the same at reference frequency, which is at the very beginning of the signal, and not at the respective segment start times.

Further, assuming a common $\theta_{JN}$ or inclination $\iota$ does not restrict the angular momentum direction of the binary. It assumes only that the observer views the binary at the same angle in both parts of the segment. Any deviations in the total angular momentum of the signal at the reference frequency, between the inspiral and ringdown signals, would be captured in the exclusive parameters of the system, including the spin vectors and the derived orbital angular momentum vector. 

Thus, such a segregation of the 15-dimensional BBH parameter space into exclusive and common sets of sampled parameters allows the joint determination of the common parameters based on combined information available across the signal segments, while at the same time allowing correlations with their respective exclusive parameters to be different. This, therefore, forces deviations from GR to be captured in the non-geometric parameters (which we identify with the exclusive parameters) of the system. 

Note that having no common set of parameters, i.e., $\thetabf_c \equiv \emptyset $, amounts to fully independent analysis, as has always been carried out before, usually in the frequency domain, and in the time domain in~\cite{LIGOScientific:2025rid}. Due to parameter correlations, independent analyses can yield slightly different estimates of the source's geometric location and orientation, leading to an inconsistent source picture and complicating comparisons due to the additional degeneracies. Keeping them common ensures that the geometric configuration of the source is jointly determined by pooling the relevant extrinsic information, while preserving their respective correlations with the other exclusive parameters.

However, if there is reason to question/test these assumptions, or to artificially project certain deviations onto the geometric variables, the framework provides the user with the flexibility to choose the system's common parameters at the start of the analysis.
\subsection{Injection test}
\label{sec:inj}
%

I perform several injection studies to test the robustness of the procedure. For concreteness, we discuss one such injection, carried out using the maximum likelihood parameters from a time-domain PE of GW250114. To demonstrate the capability of the MSCT, we set the inspiral analysis to end at about $190M$ before the invariant peak of the polarization in GPS time. The median SNR of the inspiral signal is about $31$. The ringdown segment is set to begin at $\sim 28M$ after the peak of the polarization, and its SNR is only about $7$. The combined SNR of the analyzed data is approximately $32$.
\begin{table}[t]
\centering
\renewcommand{\arraystretch}{1.5}
\begin{tabular}{l c}
\toprule
Parameter & Median ($^{+\,\mathrm{upper}}_{-\,\mathrm{lower}}$) \\
\midrule
\multicolumn{2}{l}{\textbf{Differences}}\\
$\Delta \chi_{\mathrm{eff}}$ & $6.44\times10^{-3}{}^{+0.59}_{-0.52}$ \\
$\Delta \chi_{\mathrm{p}}$   & $-5.16\times10^{-2}{}^{+0.4}_{-0.38}$ \\
$\Delta \mathcal{M}$         & $1.76\times10^{-2}{}^{+5.59}_{-4.26}M_\odot$ \\
$\Delta q$                   & $0.134{}^{+0.31}_{-0.32}$ \\
$\Delta \phi$                & $-0.950{}^{+3.96}_{-3.14}$ \\
$\Delta \psi$                & $-2.21\times10^{-2}{}^{+1.8}_{-1.85}$ \\
$\Delta \phi_{12}$           & $-6.87\times10^{-2}{}^{+3.24}_{-3.5}$ \\
$\Delta \phi_{JL}$           & $-0.45{}^{+4.03}_{-4.22}$ \\
$\Delta \theta_{1}$          & $0.2{}^{+1.44}_{-1.35}$ \\
$\Delta \theta_{2}$          & $-0.32{}^{+1.26}_{-1.32}$ \\
$\Delta M_f$                 & $1.86{}^{+10.38}_{-8.73}M_\odot$  \\
$\Delta \chi_f$              & $3.17\times10^{-3}{}^{+0.19}_{-0.22}$ \\
\end{tabular}
\caption{Medians and the 10, 90\% quantiles on the marginal distribution for the deviation parameters from across the segments, for the zero-noise injection of a GW250114-like signal.}
\label{tab:diff_pars_inj}
\end{table}

\end{document}